\documentclass[a4paper, amsfonts, amssymb, amsmath, reprint, showkeys, twoside,superscript address]{revtex4-2}

\newcommand{\mez}{\hspace*{+0.50cm}}
\newcommand{\mz}{\hspace*{+0.25cm}}
\newcommand{\m}{\hspace*{-0.50mm}}
\newcommand{\n}{\hspace*{-0.25mm}}
\newcommand{\be}{\begin{equation}}
\newcommand{\ee}{\end{equation}}
\newcommand{\la}{\langle}
\newcommand{\ra}{\rangle}
\usepackage{bm,amssymb,amsmath}
\usepackage{graphicx,color,upgreek}
\usepackage{float}
\usepackage{xcolor}
\usepackage{relsize}
\usepackage{comment}

\begin{document}

\vspace*{+1.00cm}

 \title{Steady state  of periodically driven quantum systems}
 \author{ Milan \v{S}indelka}
 \email{\tt milan.sindelka@gtiit.edu.cn}
 \affiliation{Guangdong Technion Israel Institute of Technology,
241 Daxue Road, Shantou, Guangdong Province 515603, P.R. China}
 \affiliation{Schulich Faculty of Chemistry and Helen Diller Quantum Center, Technion -- Israel Institute of Technology, Haifa 32000, Israel}
 \author{David Gelbwaser-Klimovsky}
 \email{\tt dgelbi@technion.ac.il}
 \affiliation{Schulich Faculty of Chemistry and Helen Diller Quantum Center, Technion-Israel Institute of Technology, Haifa 3200003, Israel}

\begin{abstract}
Periodic driving is used to steer physical systems to unique stationary states or nonequilibrium steady states (NESS), producing enhanced properties inaccessible to non-driven systems. For open quantum systems, characterizing the NESS is challenging and existing
results are generally limited to specific types of driving and the Born-Markov approximation. Here we go beyond these limits by studying a generic periodically driven $ N$-level quantum system interacting with a low-density thermal gas. Exploiting the framework of
 Floquet scattering theory, we establish general Floquet thermalization conditions constraining the nature of the NESS and the transition rates. Moreover, we examine theoretically the structure of the NESS  in the high temperature limit,
 and find out that the NESS complies, rather surprisingly,  with
 an uniform probability distribution (predicted by the Boltzmann law) for any driving. Numerical calculations illustrate our theoretical elaborations for a simple toy model.
\end{abstract}

\maketitle

\section{Introduction}
\mez Periodic driving is used in a wide range of fields to modify quantum systems  in an ad hoc manner and to enhance  specific properties such as:  tunneling suppresion in BEC \cite{lignier2007dynamical}, inducing  topological state in semiconductor quantum wells \cite{lindner2011floquet} and creating topological and a quantum spin Hall insulator in lattices \cite{hauke2012non}, reducing decoherence of a spin-$\frac{1}{2}$ particle \cite{chen2015floquet}, creating quantum heat machines \cite{gelbwaser2013minimal}. This approach is known as Floquet engineering and has been studied both for closed \cite{bukov2015universal,eckardt2017colloquium} and open systems \cite{mori2023floquet}.

\mez Periodically driven systems reach a periodic steady state or limiting cycle \cite{yudin2016dynamic}.
For  systems interacting with a thermal bath, the periodic steady state can be mapped to a nonequilibrium steady state
(NESS) \cite{raz2016mimicking}. The NESS ultimately determines the steady physical properties of the system. In general, determining the NESS is a complex problem, but some progress has been made under some restrictions on the driving. For example,  in \cite{ikeda2020general,ikeda2021nonequilibrium,martins2025quasiperiodic},
the NESS of a high-frequency periodically driven system was found. To avoid the complexity of finding the NESS, an alternative approach can be taken: studying the conditions under which the NESS does not deviate from the thermal state or a thermal state with an effective Hamiltonian, also called the Floquet-Gibbs state (FGS). In these cases, all the novel properties or phases inaccessible in thermal equilibrium are lost, and therefore, these conditions should be avoided if  any enhanced properties are desired. In particular, under the Born-Markov approximation or weak coupling limit \cite{breuer2002theory},
detailed balance of the transition rates has been used to find conditions for a FGS \cite{breuer2000quasistationary,shirai2016effective,liu2015classification}. These conditions are limited to specific driving Hamiltonians or frequencies, and, therefore, can not be universally applied.

\mez In this work, we study the NESS of a periodically driven $N$-level system interacting with a single thermal bath represented by a dilute gas. This allows us to 
 employ a Lindblad equation beyond the weak coupling limit \cite{dumcke_low_1985},  to study the NESS beyond the Born-Markov approximation, and to find conditions for it to be a FGS. Exploiting the framework of Floquet scattering theory
\cite{peskin1994time,lefebvre2005scattering}, we establish general Floquet thermalization conditions constraining the
nature of the NESS and the transition rates. These conditions generalize the usual detailed balance-like restriction commonly found for rates of periodically driven system under the Born-Markov approximation \cite{kohn2001periodic}. Moreover, we examine theoretically the structure of NESS at high temperatures, and
find out that the high temperature NESS  approaches, rather surprisingly,  to the
uniformly distributed FGS, independently of the driving strength, frequency or Hamiltonian. This happens despite the lack of compliance with detailed balance. As we show, the lack of deviation from a Boltzmann state
in the high temperature regime
is a consequence of the unitarity of the scattering  operator,  which at high temperatures precludes the rate asymmetries required for a non-thermal steady state. In contrast, for lower temperatures, the NESS deviates from the Boltzmann equilibrium state, requiring the injection of energy in order to produce it. Our theoretical elaborations are illustrated by numerical calculations for a simple toy model.

\section{Theoretical outline}

\subsection{Periodically driven $\bm{N}$-level system coupled to dilute thermal gas}

\mez We consider an $ N$-level quantum system that is subjected to a periodic driving with frequency $\omega$. Moreover, the quantum system interacts with a single thermal bath at an inverse temperature $\beta=(k_BT)^{-1}$. We furthermore assume that the bath is a $d$-dimensional low-density gas  composed of free particles that are inelastically scattered (without being created or annihilated) by the driven $N$-level system. This assumption allows to write the low density Lindblad master equation for the system evolution \cite{dumcke_low_1985}, (see  S.7 in the SI for details \cite{sup}).  Such Lindblad master equation is valid for any bath -- system coupling strength even beyond the Born-Markov approximation.

\mez Periodic driving is described by exploiting the language of Floquet theory \cite{peskin1994time,lefebvre2005scattering} (see S.1  in the SI \cite{sup}).
Due to driving, bare energy levels $\{ E_j, 1 \leq j \leq N \}$ of the $N$-level system are converted into quasi-energies
$\{ E_{j\nu}^{\rm QE} = E_{j0}^{\rm QE} + \nu\hbar\omega, 1 \leq j \leq N, \nu \in {\mathbb Z} \}$, where $E_{j0}^{\rm QE}$
reduces to $E_j$ for zero driving, and the Floquet index $\nu$
labels here different Brillouin copies of the same physical Floquet state $j$.

\mez
Hereafter we shall assume non-degenerate setup, $E_{j\nu}^{\rm QE} \neq E_{j'\nu'}^{\rm QE}$ for $(j\nu) \neq (j'\nu')$. In such non-degenerate cases, the
 physical populations $\wp_{j}$ of the driven $N$-level system are decoupled from coherences, and the latter decay to zero.
From now on, we  focus solely on the populations $\wp_{j}$  of the physical Floquet states $j$,
which evolve  according to the Pauli rate equation \cite{pauli1928sommerfeld}: 

\begin{eqnarray} \label{Pauli}
 \dot{\wp}_{j} & = & \sum_{j'\n=1}^{N} \, a_{jj'} \, \wp_{j'} \; - \; \wp_{j} \, \sum_{j'\n=1}^{N} \, a_{j'\n j} \;\; ;
\end{eqnarray}
with
$a_{j'\n j} \; = \; \sum_{\nu=-\infty}^{\nu=+\infty} a_{j'\n j}^\nu$.
Here  $a_{j'\n j}^{\nu}$ refers to the rate of transition $j \to j'$ and the absorption  of $\nu$ Floquet quanta  by the system from the driving field. These transition rates are obtained using the framework of the low-density limit Lindblad equation. They are based on scattering theory and are  constructed from the Floquet $\widehat{T}-$matrix   elements through the following formula,
 \begin{eqnarray} \label{Floquet-rates}
 & & a_{j'\n j}^{\nu} \; = {\cal N} \, Z^{-1} \m \int_{{\mathbb R}^d} \m {\rm d}^dp \int_{{\mathbb R}^d} \m {\rm d}^dp' \; e^{-\beta\frac{\bm{p}^2}{2\,m}} \, \times \\
 &&
 \delta\m\m\left(\frac{\bm{p}'^2}{2\,m}+E_{j'\n\nu}^{\rm QE}-\frac{\bm{p}^2}{2\,m}-E_{j0}^{\rm QE}\right) \,
 \Bigl| \la\m\n\la \bm{p}'\n\,(j'\n\nu) |\n| \widehat{T} |\n| \bm{p}\,(j0) \ra\m\n\ra \Bigr|^2 , \nonumber
 \end{eqnarray}
see (S3) and S.7 in the SI  \cite{sup}.
Here the prefactor ${\cal N}$ is proportional to the gas particle density, $Z=\int_{{\mathbb R}^d} \n {\rm d}^dp \, e^{-\beta\frac{\bm{p}^2}{2\,m}}$, and $\la\m\n\la \bm{p}'\n\,(j'\n\nu) |\n| \widehat{T} |\n| \bm{p}\,(j0) \ra\m\n\ra$ are the on-shell
Floquet $\widehat{T}$-matrix elements, encoding all information about scattering of a single gas particle on the driven $N$-level system
(see S.1  in the SI  \cite{sup}).  Vector
 $|\n| \bm{p}\,(j0) \ra\m\n\ra$ represents an asymptotic in-state corresponding to a gas particle of momentum $\bm{p}$ which approaches the driven system residing in its in-Floquet state $(j0)$. Similarly $\la\m\n\la \bm{p}'\n\,(j'\n\nu) |\n|$ refers to  an asymptotic out-state corresponding to a gas particle of momentum $\bm{p}'$ which departs from the driven system residing in its out-Floquet state $(j'\n\nu)$.

For $\dot{\wp}_j=0$, we obtain from (\ref{Pauli}) the NESS, $\wp^{\rm NESS}_j$,
of the driven $N$-level system. The corresponding stationarity condition,
\begin{eqnarray} \label{NESS-def}
 \sum_{j'\n=1}^{N} \, a_{jj'} \, \wp^{\rm NESS}_{j'} \; - \; \wp^{\rm NESS}_{j} \, \sum_{j'\n=1}^{N} \, a_{j'\n j} \; = \; 0 \;\; ,
\end{eqnarray}
together with  normalization requirement
$\sum_j \wp^{\rm NESS}_j = 1$, determine the NESS $\{ \wp^{\rm NESS}_j \}_{j=1}^{N}$ uniquely in terms of
the total rates $\{ a_{jj'} \}$.

\subsection{Floquet thermalization conditions}

\mez  We recall that the  Floquet $\widehat{S}$-matrix elements take the form
\begin{eqnarray} \label{widehat-S-Floquet}
 && \la\m\n\la \bm{p}'\n\,(j'\n\nu) |\n| \widehat{S} |\n| \bm{p}\,(j0) \ra\m\n\ra = \delta^d\m(\bm{p}'\n-\bm{p}) \,
 \delta_{j'\n j} \, \delta_{\nu0} -\\  & 2&\,\pi\,i \; \delta\m\left( \frac{\bm{p}'^2}{2\,m} + E_{j'\n\nu}^{\rm QE}
 - \frac{\bm{p}^2}{2\,m} - E_{j0}^{\rm QE} \right) \, \la\m\n\la \bm{p}'\n\,(j'\n\nu) |\n| \widehat{T} |\n| \bm{p}\,(j0) \ra\m\n\ra \;
 ; \nonumber
\end{eqnarray}
see S.2  in the SI  \cite{sup} and Refs.~\cite{peskin1994time,lefebvre2005scattering}. All scattering processes preserve unitarity,
due to the unitary character of the quantum dynamical evolution. The resulting unitarity of $\widehat{S}$ imposes in turn strong constraints on the associated $\widehat{T}$-matrix elements.  Such constraints are also known as the optical theorem (see Eq.~(36) in
Ref.~\cite{takayanagi2023generalized} and Eqs.~(S5)-(S16) in the  SI  \cite{sup}).
The just mentioned optical theorem can be used
to deduce constraints on the  Floquet transition rates $a_{j'\n j}^{\nu}$. Namely, one finds that
 \be \label{thermalization-Floquet-general-take-2-prelim}
 \hspace*{-0.25cm}
 \sum_{j'\n=1}^{N} \sum_{\nu'\n=-\infty}^{\nu\n=+\infty} \, a_{jj'}^{-\nu} \, e^{-\beta E_{j'\n\nu}^{\rm QE}} \; = \;
 e^{-\beta E_{j0}^{\rm QE}} \, \sum_{j'\n=1}^N \sum_{\nu\n=-\infty}^{\nu'\n=+\infty} a_{j'\n j}^{+\nu} \;\; ;
 \ee
valid for all $1 \leq j \leq N$ and $\nu \in {\mathbb Z}$ (see  S.2 in the SI  \cite{sup}).

\begin{figure}[h!]
\includegraphics[scale=0.6,angle=0]{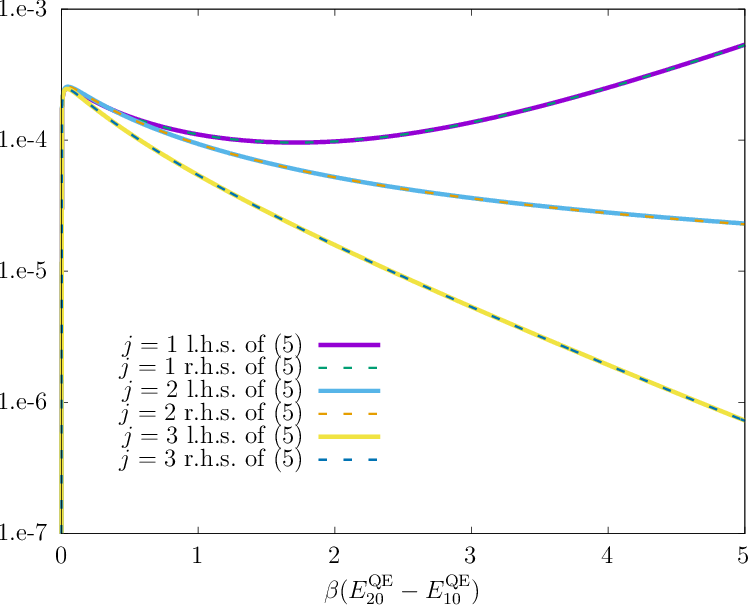}
\caption{Numerical check of the Floquet thermalization conditions (\ref{thermalization-Floquet-general-take-2-prelim}) for a driven three-level system. 
 Details of the used model can be found  in section III of the main text and in  S.6 of the SI. \label{fig:thcond}}
\end{figure}

\mez The constraints (\ref{thermalization-Floquet-general-take-2-prelim})
have important thermodynamic consequences for the nature of the NESS, as shown below.
Conditions (\ref{thermalization-Floquet-general-take-2-prelim}) are less restrictive than the standard detailed balance-like relations generally used in periodically driven open quantum systems: $ a_{jj'}^{-\nu} \, e^{-\beta E_{j'\n\nu}^{\rm QE}} =
 e^{-\beta E_{j0}^{\rm QE}}  a_{j'\n j}^{+\nu}$ \cite{kohn2001periodic}. The  just mentioned detailed balance-like relations are valid only in the Born-Markov approximation or for systems that do not break microreversibility \cite{alicki2023violation} (that is, in the absence of a magnetic field or any other velocity dependent forces). Because our  dynamical evolution equation \eqref{Pauli} reaches beyond the Born-Markov approximation,  we will use the more general  condition~(\ref{thermalization-Floquet-general-take-2-prelim}). Nevertheless, our results are also valid for regimes where the detailed balance-like relations  apply.

 \mez For
 zero driving, 
 $a_{j'\n j}^{\nu} = \delta_{\nu0} \, a_{j'\n j}$. Accordingly, relations
 (\ref{thermalization-Floquet-general-take-2-prelim}) boil down into
 \be \label{thermalization-Floquet-general-take-2-nondriven}
 \sum_{j'\n \neq j} \, a_{jj'} \, e^{-\beta E_{j'}} \; = \;
 e^{-\beta E_{j}} \sum_{j'\n \neq j} a_{j'\n j} \;\; ;
 \ee
for all $1 \leq j \leq N$. Properties (\ref{thermalization-Floquet-general-take-2-nondriven}) represent nothing else than the
 thermalization conditions encountered in the theory of non-driven open quantum systems \cite{alicki2023violation,blum_thermalization_2025}.
 We have thus shown that unitarity of the total (system + bath) dynamics implies the thermalization of a non-driven system that undergoes a non-unitary reduced dynamics. This represents a mechanism of emergence of thermalization of a reduced system under a total unitary dynamics. Its connection with other mechanisms such as the eigenstate thermalization hypothesis (ETH) \cite{o2025quantum,d2016quantum} should be further explored. An analogy between (\ref{thermalization-Floquet-general-take-2-nondriven}) and
(\ref{thermalization-Floquet-general-take-2-prelim}) motivates us to hereafter call our more
general relations (\ref{thermalization-Floquet-general-take-2-prelim}) as the Floquet
thermalization conditions.
We note that constraints  (\ref{thermalization-Floquet-general-take-2-prelim})
 or (\ref{thermalization-Floquet-general-take-2-nondriven})
are
 fundamental
preconditions that should be fulfilled by the rates of the Pauli dynamics of any physical $N$-level system coupled to a dilute thermal gas.

\subsection{Structure of the NESS}

\mez The Floquet thermalization conditions (\ref{thermalization-Floquet-general-take-2-prelim}) can also be redisplayed as follows:
\be \label{thermalization-Floquet-general-v1}
 \sum_{j'} \, a_{jj'} \, e^{-\beta E^{\rm QE}_{j'0}} \m \left\la e^{\beta\nu\hbar\omega} \right\ra_{\m jj'}
 \; = \; e^{-\beta E^{\rm QE}_{j0}} \, \sum_{j'} \, a_{j'\n j} \;\; ,
\ee
where
\begin{equation}
\label{exponential-factors}
 \left\la e^{\beta\nu\hbar\omega} \right\ra_{\m jj'}  =
  a_{jj'}^{-1}  \sum_{\nu}
 \, a_{jj'}^{\nu} \, e^{\beta\nu\hbar\omega}  =
 \frac{\sum_{\nu} a_{jj'}^{\nu} \, e^{\beta\nu\hbar\omega}}{\sum_{\nu} a_{jj'}^{\nu}} \;\; .
\end{equation}
The term
  $\left\la e^{\beta\nu\hbar\omega} \right\ra_{\m jj'}$ represents a measure of how much the NESS deviates from the thermal state due to driving.
 This  becomes clearer  after using (\ref{thermalization-Floquet-general-v1}) and rewriting Eq.~\eqref{NESS-def} as
\begin{equation}
\hspace*{-0.25cm}
\sum_{j'}a_{jj'}\left(\frac{\wp^{\rm NESS}_{j'}}{\wp^{\rm NESS}_{j}}-e^{-\beta({ E}^{\rm QE}_{j'\n 0}-{ E}^{\rm QE}_{j0})}\langle e^{\beta\nu\hbar\omega}\rangle_{jj'}\right)=0 \; . \label{eq:ss}
\end{equation}
If $\left\la e^{\beta\nu\hbar\omega} \right\ra_{\m jj'}=1$ for all $j$ and $j'$,  equation \eqref{eq:ss} implies that the steady state is thermal. This is the case for the non-driven case,  and also for the high temperature limit where $\left\la e^{\beta\nu\hbar\omega} \right\ra_{\m jj'}\rightarrow 1$. Therefore, at infinite temperature, the steady state is thermal  and has thus uniform population distribution, $\wp^{\rm NESS}_{j}=1/N$. This uniformity can also be deduced from the Floquet thermalization conditions. Namely, for $\beta=0$, Eq.~(\ref{thermalization-Floquet-general-v1}) implies that the transition rates are symmetric in the sense that the sum of transition rates of the processes flowing out of state $j$ is balanced by the sum of transition rates of processes going into state $j$.  Such symmetry results in a steady state with a homogeneous distribution.
\begin{figure}[h!]
\includegraphics[scale=0.6,angle=0]{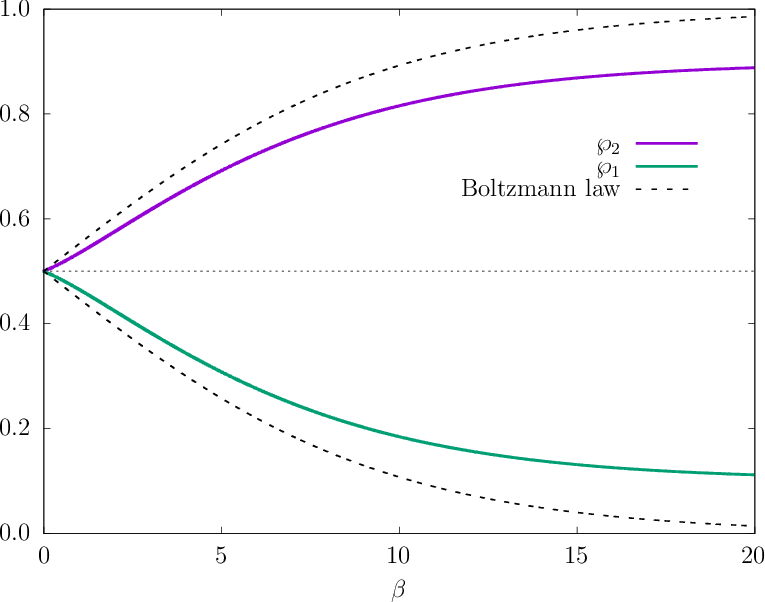} 
\caption{ The structure of NESS for a simple toy model of a driven two-level system. The uniform population distribution emerges at $\beta=0$.
Details of the used model can be found  in section III of the main text and in  S.6 of the SI.
} 
\end{figure}

\mez For a thermal steady state, on average no energy is exchanged with either the bath or the driving.
The energy exchange with the driving is comprised of multiple processes, each involving a specific number of Floquet excitations, $\nu$. The probability of each  such process is the product between the state population, $\wp^{\rm NESS}_{j}$\m, and the transition rate $a_{jj'}^{\nu}$. Using  the fact that at infinite temperature the steady state is proportional to the identity, i.e.,
  it is a thermal state, a zero energy exchange with the driving imposes a further constraint on the transition rates at
  $\beta=0$. Namely,

\begin{equation} \label{eq:nocht-prelim}
 \hbar\omega \sum_{j} \sum_{j'} \,\sum_{\nu=-\infty}^{\nu=+\infty}  \nu \, a_{jj'}^{\nu} \, \Biggr|_{\beta=0}=0 \;\; ; 
\end{equation}
or, equivalently,
\begin{eqnarray} \label{eq:nocht}
 & &  \hbar\omega \sum_{j} \sum_{\nu=-\infty}^{\nu=+\infty}  \nu \, a_{jj}^{\nu} \, \Biggr|_{\beta=0}  \\
 & + & \hbar\omega \sum_{j > j'} \,\sum_{\nu=-\infty}^{\nu=+\infty}  \nu \, (a_{jj'}^{\nu}+a_{j'j}^{\nu}) \, \Biggr|_{\beta=0}=0 \;\; . \nonumber
\end{eqnarray}

\mez Our  just given thermodynamic argument makes the result (\ref{eq:nocht-prelim}) independent of the driving strength, frequency and the Hamiltonian, and therefore it can be universally applied to any driving. Eq.~(\ref{eq:nocht-prelim}) or (\ref{eq:nocht}) can also be derived  in an
alternative manner not relying on thermodynamics, as a mere consequence of the Floquet thermalization conditions  (see S.3 in the SI  \cite{sup}).
\begin{figure}[h!]
\includegraphics[scale=0.6,angle=0]{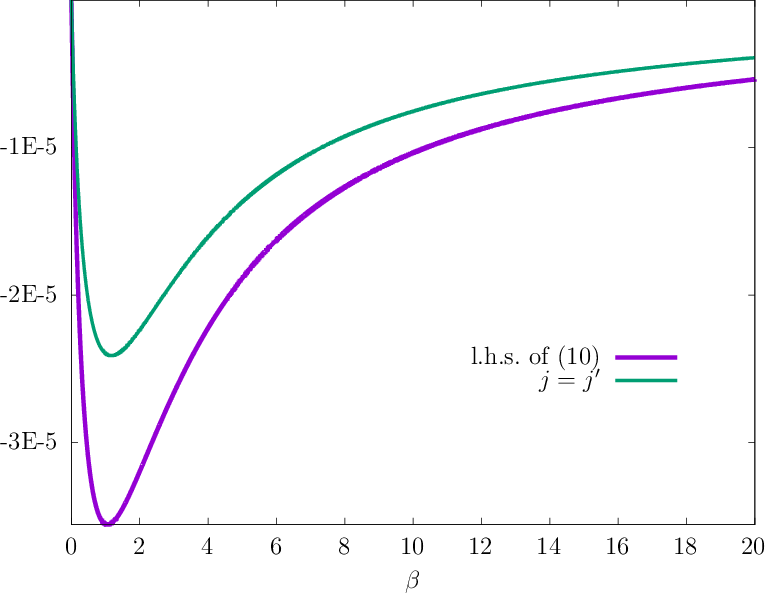} 
\caption{ 
Quantity $\hbar\omega \sum_{j} \sum_{j'} \,\sum_{\nu=-\infty}^{\nu=+\infty}  \nu \, a_{jj'}^{\nu}$ plotted against $\beta$
for a simple toy model of a driven two-level system. One can see that the property (\ref{eq:nocht-prelim}) holds at $\beta=0$.
Interestingly, the green line shows that even $\hbar\omega \sum_{j} \sum_{\nu=-\infty}^{\nu=+\infty}  \nu \, a_{jj}^{\nu}$
vanishes at $\beta=0$, rationalization of such behavior is left as an open question for possible further research.
Details of the used model can be found  in section III of the main text and in S.6 of the SI.
}
\end{figure}

\mez Based on Eq.~\eqref{eq:nocht}, we conjecture that
\be \label{conjecture-2}
  (a_{jj'}^{+\nu}+a_{j'j}^{+\nu})\Bigr|_{\beta=0} \, = \, (a_{jj'}^{-\nu}+a_{j'j}^{-\nu})\Bigr|_{\beta=0} \;\; ;
\ee
otherwise fine tuning will be required to comply with Eq.~\eqref{eq:nocht}.
 The just presented conjecture can actually be proven to hold for a relatively broad class of situations.
Namely, assuming that the system -- particle coupling $\widehat{V}$ provides the matrix element $\la\m\n\la \bm{p}'\n\,(j'\n\nu) |\n| \widehat{V} |\n| \bm{p}\,(j0) \ra\m\n\ra$
which does not fall off to zero too quickly with energy (see  S.3 in the SI for a quantitative criterion  \cite{sup}), one
encounters an extra symmetry property
\be \label{Born-extra-symmetry}
   a_{jj'}^{+\nu} \, \Bigr|_{\beta=0} \, = \, a_{j'j}^{-\nu} \, \Bigr|_{\beta=0} \;\; ;
\ee
which is even much stronger than 
(\ref{conjecture-2}). Relation (\ref{Born-extra-symmetry}) is implied by
incorporating into (\ref{Floquet-rates}) the Born approximation $\la\m\n\la \bm{p}'\n\,(j'\n\nu) |\n| \widehat{T} |\n| \bm{p}\,(j0) \ra\m\n\ra \mapsto \la\m\n\la \bm{p}'\n\,(j'\n\nu) |\n| \widehat{V} |\n| \bm{p}\,(j0) \ra\m\n\ra$, such a step becomes exact for the above stated kinds of
$\widehat{V}$ (see  S.3 in the SI for details  \cite{sup}).

\mez Moreover, if extra symmetries on the driving are considered (see  S.3 in the SI  \cite{sup}), it is possible to get yet another additional strong symmetry property of the rates, namely,
\be \label{rates-extra-symmetry}
   a_{jj'}^{+\nu} \, \Bigr|_{\beta=0} \, = \, a_{jj'}^{-\nu} \, \Bigr|_{\beta=0} \;\; .
\ee
In this case, Eq.~(\ref{eq:nocht}) has  important physical consequences.  Namely, it implies that the
NESS
is thermal also for small values of $\beta$, that is,  for high but finite temperatures up to ${\cal O}(\beta)$.
\begin{figure}[h!]
\includegraphics[scale=0.6,angle=0]{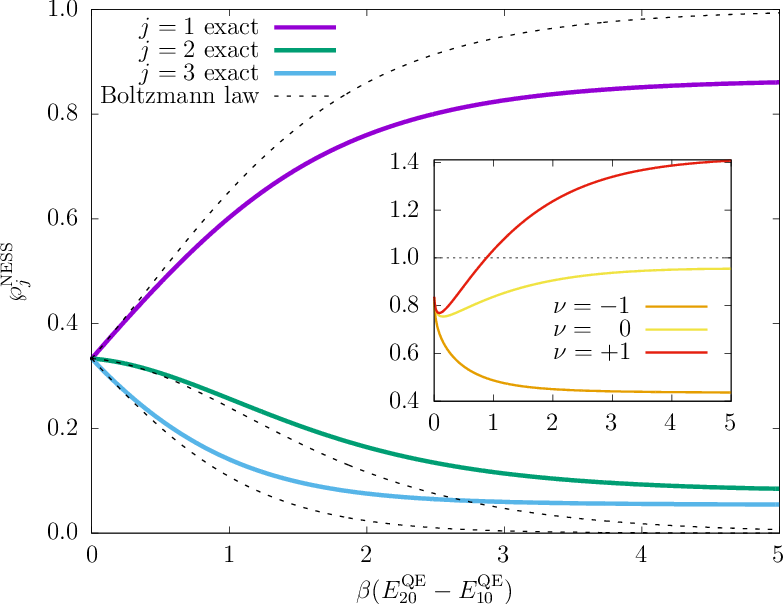} 
\caption{  The structure of NESS plotted as a function of normalized inverse temperature
$\beta(E_{20}^{\rm QE}-E_{10}^{\rm QE})$. Note the special behavior of the populations $\wp_j^{\rm NESS}$
in the $\beta \to 0$ and $\beta \to \infty$ limits.
{\sl Inset:}
$
  \left(a_{jj'}^{-\nu} \, e^{-\beta E_{j'\n\nu}^{\rm QE}}\right)\m\mathlarger{\mathlarger{\mathlarger{\mathlarger{\mathlarger{/}}}}}\m\m\left(e^{-\beta E_{j0}^{\rm QE}}  a_{j'\n j}^{\nu}\right)
$
plotted here against $\beta(E_{20}^{\rm QE}-E_{10}^{\rm QE})$ for the case of
$j=1$ and $j'=2$. Showing explicitly that the detailed balance condition does not hold.
Specifications of the used model
can be found in section III of the main text and  in S.6 of the SI.
 Note that the property (\ref{rates-extra-symmetry}) does apply for the present case, ensuring compliance with the Boltzmann law up to
${\cal O}(\beta)$.
} \label{fig:ss}
\end{figure}
To demonstrate this, we  carry out a Taylor expansion of Eq.~\eqref{eq:ss} around $\beta=0$ (see  S.5 in the SI  \cite{sup}).
Notice that the first order  expansion term of $\la e^{\beta\nu\hbar\omega} \ra_{\m jj'}$ is proportional to the l.h.s.~of Eq.~\eqref{eq:nocht}. Defining $\wp^{\rm corr}_{j'\n j}$ as the first order correction to the steady state population rates, that is, $\frac{\wp^{\rm NESS}_{j'}}{\wp^{\rm NESS}_{j}}=1+\beta \wp^{\rm corr}_{j'\n j}+ O(\beta^2)$, we get
\begin{eqnarray}
\wp^{\rm corr}_{j'\n j} & = &
-({ E}^{\rm QE}_{j'0}-{ E}^{\rm QE}_{j0}) \; + \; \frac{\sum_{\nu}\nu \, a_{j'\n j}^{\nu}|_{\beta=0}}{\sum_{\nu} a_{j'\n j}^{\nu}|_{\beta=0}} \; =\notag\\ & = & -({ E}^{\rm QE}_{j'0}-{ E}^{\rm QE}_{j0}) \;\; ; \label{eq:pcor}
\end{eqnarray}
where we used the fact that no energy is exchanged with the driving  field at $\beta=0$, i.e., Eq.~\eqref{eq:nocht}, to obtain the second equality. The r.h.s.~of Eq.~\eqref{eq:pcor} is nothing but the first order correction to the thermal state. Therefore,  the NESS equals to the thermal state up to ${\cal O}(\beta)$ (see Fig.~\ref{fig:ss}).

\mez The fact that the steady state approaches the thermal state at high tempeatures has important implications for any practical realization of a periodically driven system, because the strength of the driving is always limited. Once the maximal driving strength is determined, one should be careful to avoid working at too high temperatures, otherwise the steady state will be thermal.

\mez
The precise meaning of ``too high temperatures" depends  upon whether the driving complies with  the extra symmetry property (\ref{rates-extra-symmetry}) or not. If it does not, deviations from the thermal state start at first order  of $\beta$
and high temperatures are defined as
\begin{equation} \label{david-criterion1}
 \beta \; \ll \; \frac{1}{-({ E}^{\rm QE}_{j'0}-{ E}^{\rm QE}_{j0}) \; +\frac{ d \left\la e^{\beta\nu\hbar\omega} \right\ra_{\m jj'}}{d\beta}\Bigr|_{\beta=0}} \; \; .
\end{equation}

 On the other hand, if the condition (\ref{rates-extra-symmetry}) is matched, the thermal state deviations start only at the second order  of $\beta$, and the creation of non-thermal states could be more challenging.
  In this case, the notion of high temperature can be estimated by calculating the second order $\beta$-expansion of Eq.~(\ref{eq:ss}) and comparing it to the first  order correction. If
\begin{equation} \label{david-criterion}
 \beta \; \ll \; \frac{2\,\left({ E}^{\rm QE}_{j'0}-{ E}^{\rm QE}_{j0}\right)}{\frac{ d^2 \left\la e^{\beta\nu\hbar\omega} \right\ra_{\m jj'}}{d\beta^2}\Bigr|_{\beta=0}+\left({ E}^{\rm QE}_{j'0}-{ E}^{\rm QE}_{j0}\right)^{\m\n 2}} \;\;\; ;
\end{equation}
then the NESS is well approximated by a thermal state (see Fig.~4).
The effects of driving on determining the high temperature  regime are captured here by
$\frac{ d^2 \left\la e^{\beta\nu\hbar\omega} \right\ra_{\m jj'}}{d\beta^2}\Bigr|_{\beta=0}$.
\begin{figure}[h!]
\includegraphics[scale=0.6,angle=0]{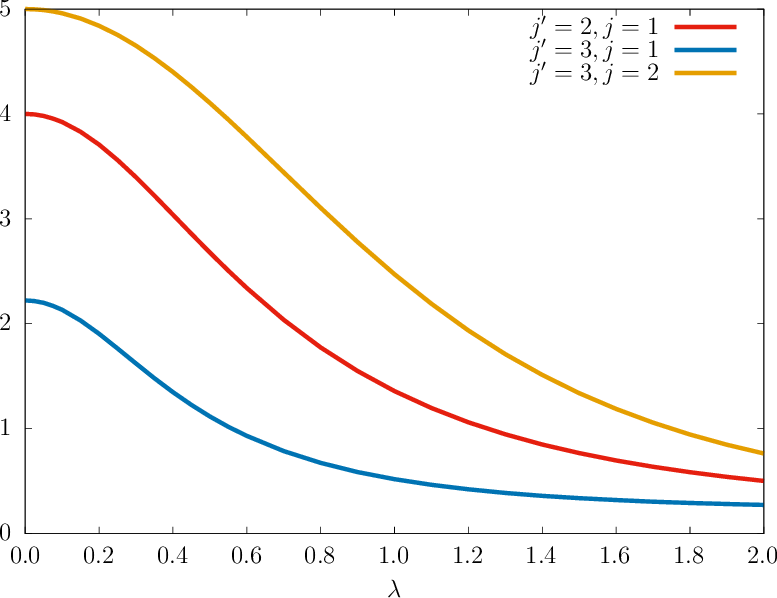} 
\caption{ High temperature limit (r.h.s.~of (\ref{david-criterion})).
Here,  the $\lambda$-dependence saturates for large values of $\lambda$.
Therefore, for the specific system and type of driving used in our calculation,
the pertinent NESS is basically thermal for $\beta \ll 0.2$, regardless upon the strength of driving.
This saturation phenomenon deserves to be explored in more detail in the future. Specifications of the used  three level model
complying with (\ref{rates-extra-symmetry})
can be found in section III of the main text and  S.6 of the SI}.
  \label{fig:hight}
\end{figure}

\mez In contrast to the high temperatures behavior, at low temperature $(\beta \to \infty)$ the NESS asymptotes always deviate from the thermal state (see Figs.~2 and 4). This is analytically proven in  S.4 of the SI.

\section{Numerical test}
  \mez We have tested our 
  theoretical findings using a toy model of a periodically driven  two and
  three level system interacting with a dilute thermal gas.
   The Hamiltonian of our driven two level system takes the form
\begin{eqnarray} \label{H-S-t-TLS}
 \hat{H}_{\rm S}\n(t) & = & E_{2} \, |2\ra\la2| \; + \; E_1\,|1\ra\la1| \nonumber \\
 & + & \lambda \cos(\omega t) \, |1\ra\la2| \; + \; \lambda \cos(\omega t) \, |2\ra\la1| \;\; ;
\end{eqnarray}
this corresponds e.g.~to an atomic system driven by laser. The Hamiltonian of our driven three level system takes the form
\begin{eqnarray} \label{H-S-t}
 \hat{H}_{\rm S}\n(t) & = & \Big( E_{3}+\lambda \cos(\omega t) \Big) \, |3\ra\la3| \; + \; E_2\,|2\ra\la2| \nonumber \\
 & + & \Big( E_{1}-\lambda \cos(\omega t) \Big) \, |1\ra\la1| \;\; .
\end{eqnarray}
In equations (\ref{H-S-t-TLS}) and (\ref{H-S-t}),
$\omega$ represents the driving frequency, and $\lambda$ controls the driving strength. The interaction with
the low density thermal gas is described by a repulsive localized potential that acts on the gas particle and a system operator that is not diagonal in the eigenbasis of $\hat{H}_{\rm S}\n(t)$ (for details, see S.6 in the SI  \cite{sup}). The corresponding $\widehat{T}$-matrix elements can be numerically calculated as detailed in S.6 of the SI, where  also all values of the parameters of the two models are included.
Using Eq.~\eqref{Floquet-rates} the transition rates can finally be obtained. Hereafter we shall adopt the following system parameters:
$E_1=-0.25$, $E_2=+0.25$, $\omega=0.7$, $\lambda=0.5$ (two level system), and $E_1=-0.5$, $E_2=0.0$, $E_3=0.4$,
$\omega=1.35$, $\lambda=0.5$ (three level system, in Fig.~5 the parameter $\lambda$ varies continuously).

\mez Our above presented analytic formulas also offer potentially important
criteria for testing numerical calculations.  In particular, the Floquet thermalization conditions
(\ref{thermalization-Floquet-general-take-2-prelim}) and the constraints (\ref{eq:nocht}) on the rates
at $\beta=0$ can be used to test the accuracy of numerical calculations of these transition rates.
The former point is illustrated in Fig.~6.
\begin{figure}[h!]
\hspace*{-0.20cm}
\includegraphics[scale=0.55,angle=0]{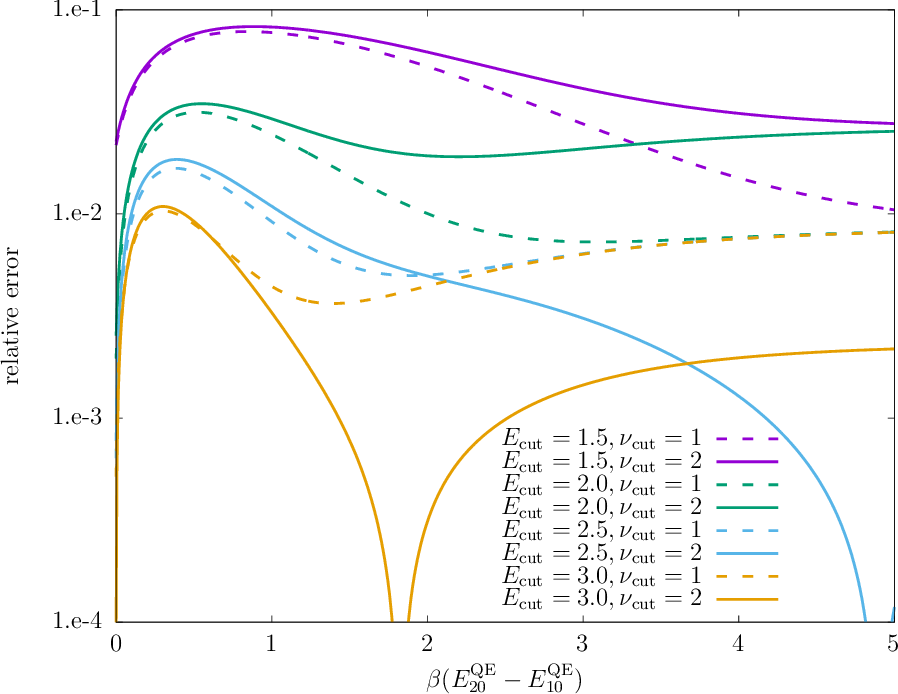}
\caption{ Floquet thermalization conditions (\ref{thermalization-Floquet-general-take-2-prelim}) used as a criterion of accuracy of the calculated transition rates, $a_{j'\n j}^\nu$. Vertical axis corresponds here to the relative error, i.e., to $2\,|({\rm l.h.s.}\,{\rm of}\,(\ref{thermalization-Floquet-general-take-2-prelim}))-({\rm r.h.s.}\,{\rm of}\,(\ref{thermalization-Floquet-general-take-2-prelim}))|/(({\rm l.h.s.}\,{\rm of}\,(\ref{thermalization-Floquet-general-take-2-prelim}))+({\rm r.h.s.}\,{\rm of}\,(\ref{thermalization-Floquet-general-take-2-prelim})))$. $E_{\rm cut}$ = cufoff energy of the incoming gas particles, $\nu_{\rm cut}$ = cutoff value of magnitude of the Floquet index $\nu$.
The relative error gradually decreases when $E_{\rm cut}$ and $\nu_{\rm cut}$ are lifted up.
Specifications of the used  three level model can be found in section III of the main text and in S.6 of the SI.
\label{fig:thcond-num}}
\end{figure}

\section{Concluding remarks}

\mez In summary, we have studied a generic periodically driven $ N$-level quantum system interacting with a low-density thermal gas.
We have found that unitarity of the scattering matrix (\ref{widehat-S-Floquet}) gives rise to the
Floquet thermalization conditions (\ref{thermalization-Floquet-general-take-2-prelim}), constraining the transition rates
$a_{j'\n j}^{\nu}$ of equation (\ref{Floquet-rates}). This forces the NESS to be close to  an uniformly distributed (thermal) state at high temperatures, and suppresses the steady energy exchange with external driving, and the steady heat flow with the bath. Therefore,  we conclude that high temperatures reduce the possibility of driving a system away from uniformly distributed thermal equilibrium. The just mentioned finding holds universally for \emph{any driving and strength of the system-bath coupling} (although the notion of high enough temperatures depends on the driving type).

\mez Moreover,  we note that our quantum mechanical derivation of the
standard thermalization conditions (\ref{thermalization-Floquet-general-take-2-nondriven}), valid for nondriven systems, represents
a novel insightful contribution of its own right, with potential implications in the field of emergence of thermalization. Finally,   conditions \eqref{thermalization-Floquet-general-take-2-prelim} also represent a new tool to test the accuracy of numerical calculations.


\mez
\textit{Acknowledgement.}
This research project was partially/fully supported by the Helen Diller Quantum Center at the Technion.
D.G.K.~is  supported by the ISRAEL SCIENCE FOUNDATION (grant No. 2247/22) and by the Council for Higher Education Support Program for Hiring Outstanding Faculty Members in Quantum Science and Technology in Research Universities.

\newpage

\onecolumngrid
\renewcommand{\theequation}{S\arabic{equation}}

\setcounter{figure}{0}
\renewcommand{\thefigure}{S\arabic{figure}}

\setcounter{section}{0}
\renewcommand{\thesection}{S\arabic{section}}

\setcounter{subsection}{0}
\renewcommand{\thesubsection}{\thesection.\alph{subsection}}

\setcounter{equation}{0}

\newpage
\begin{center}
{\huge Supplemental material}    
\end{center}

\section{Floquet scattering theory and the transition rates}

\subsection{Floquet theory for a driven system}

\mez Let us consider a quantum system which is exposed to an external time
periodic driving. Let $\hat{H}_{\rm S}\n(t)$ be its Hamiltonian.
We have $\hat{H}_{\rm s}\n(t)=\hat{H}_{\rm s}\n(t+T)$, where $T=\frac{2\pi}{\omega}$ is the period of driving,
and $\omega$ the associated frequency.

\mez Within the framework of Floquet theory, the time variable $t$ is treated as an additional degree of freedom (an additional coordinate) {\cite{peskin1994time,lefebvre2005scattering}.
Correspondingly, one introduces the Floquet Hamiltonian $\widehat{H}_{\rm F} = \hat{H}_{\rm s}\n(t) - i\hbar\,\partial_t$ acting in an extended
state space ${\cal H}_F = {\cal H}_{\rm s} \otimes {\cal H}_{\rm extra}$. Here ${\cal H}_{\rm s}$ stands for the state space of a nondriven
target, and ${\cal H}_{\rm extra} = {\rm span}\{ e^{+i \nu \omega t} \}_{\nu \in {\mathbb Z}}$ contains all $T$-periodic functions of $t$.
Elements of ${\cal H}_F$ are hereafter denoted using a double ket notation $|\n| \Psi \ra\m\n\ra$.
Floquet states of our driven target correspond to solutions of an eigenvalue problem
\be \label{Floquet-eigenproblem}
 \widehat{H}_{\rm F} \, |\n| \Psi_{j\nu} \ra\m\n\ra \; = \; E_{j\nu}^{\rm QE} \, |\n| \Psi_{j\nu} \ra\m\n\ra \;\;\; .
\ee
In equation (\ref{Floquet-eigenproblem}), index $j$ specifies the genuine quantum state of our driven system, whereas $\nu \in {\mathbb Z}$
labels all the Brillouin copies of the same physical state. One has $E_{j\nu}^{\rm QE}=E_{j0}^{\rm QE}+\nu\hbar\omega$, with the Floquet
quasi-energy $E_{j0}^{\rm QE}$ reducing to the level $E_j$ of a nondriven system in the limit of zero driving strength.

\subsection{Floquet scattering}

\mez Consider a situation when a particle of mass $m$ carrying a given momentum $\bm{p}$
approaches our driven system residing in a given Floquet state $|\n| \Psi_{j\nu} \ra\m\n\ra$.
The pertinent incoming state $|\n| \bm{p}\,(j\nu) \ra\m\n\ra$ corresponds to energy $\frac{\bm{p}^2}{2\,m}+E_{j\nu}^{\rm QE}$.
Since the time variable $t$ is treated as an additional degree of freedom within the framework of Floquet formalism, the just encountered
collision problem is treatable by conventional methods of the time independent scattering theory. In particular, net outcome of the
collision process is described by the Floquet scattering matrix
 \begin{eqnarray} \label{Floquet-T-matels-Appendix-A}
 & & \la\m\n\la \bm{p}'\n\,(j'\n\nu') |\n| \widehat{S} |\n| \bm{p}\,(j\nu) \ra\m\n\ra \; = \\
 & = &
 \delta^d\n({\bm p}'\n-\bm{p}) \, \delta_{j'\n j} \, \delta_{\nu'\n \nu}
 \; - \; 2\,\pi\,i \; \delta\m\m\left(\frac{\bm{p}'^2}{2\,m}+E_{j'\n\nu'}^{\rm QE}-\frac{\bm{p}^2}{2\,m}-E_{j\nu}^{\rm QE}\right)
 \la\m\n\la \bm{p}'\n\,(j'\n\nu') |\n| \widehat{T} |\n| \bm{p}\,(j\nu) \ra\m\n\ra \;\;\; ; \nonumber
 \end{eqnarray}
here $\la\m\n\la \bm{p}'\n\,(j'\n\nu') |\n| \widehat{T} |\n| \bm{p}\,(j\nu) \ra\m\n\ra$ are the pertinent on-shell $\widehat{T}$-matrix elements.

\subsection{Transition rates}

\mez Next, suppose that the scattering particles correspond to a low density thermal gas.
This means that the momentum distribution of the particles is given by $Z^{-1} e^{-\beta\frac{\bm{p}^2}{2\,m}}$, where $Z=\int_{{\mathbb R}^d} \m {\rm d}^dp \; e^{-\beta\frac{\bm{p}^2}{2\,m}}$.  We are interested in the reduced dynamics of the quantum system under the influence of the particle gas. This can be derived using the low density limit Lindblad equation \cite{dumcke_low_1985}, where only single particle collision events contribute. The pertinent transition rates  can be obtained from the above defined on-shell $\widehat{T}$-matrix elements by tracing out the bath degrees of freedom. That is,
 \begin{eqnarray} \label{Floquet-rates-Appendix-A}
 & & a_{j'\n j}^{\nu} \; = \\
 & = & {\cal N} \, Z^{-1} \m \int_{{\mathbb R}^d} \m {\rm d}^dp \int_{{\mathbb R}^d} \m {\rm d}^dp' \; e^{-\beta\frac{\bm{p}^2}{2\,m}} \,
 \delta\m\m\left(\frac{\bm{p}'^2}{2\,m}+E_{j'\n\nu}^{\rm QE}-\frac{\bm{p}^2}{2\,m}-E_{j0}^{\rm QE}\right) \,
 \Bigl| \la\m\n\la \bm{p}'\n\,(j'\n\nu) |\n| \widehat{T} |\n| \bm{p}\,(j 0) \ra\m\n\ra \Bigr|^2 \;\;\; ; \nonumber
 \end{eqnarray}
 where the prefactor ${\cal N}$ is proportional to the particle number density.

\section{Thermalization conditions from scattering theory}

\subsection{Starting point}
\mez In the present section S2, we present a self contained general derivation of the thermalization conditions based upon scattering theory.
Since we wish to preserve the most general setup (which may apply for any kind of multichannel scattering, not just for the case of Floquet
scattering), we depart accordingly from the notations adopted in other parts of our work. In particular, we shall refer to energy levels
${\cal E}_n$ rather than to $E_{j\nu}^{\rm QE}$, we shall also employ a single bracket notation $|\bm\cdot\ra$ rather than the
$|\n| \bm\cdot \ra\m\n\ra$ notation. It is however a straightforward self-evident matter to relate unambiguously the more general notations of
S2 to our main text and to the remaining part of this in the supplemental material.

\mez Assume that our target system possesses discrete levels $\{{\cal E}_n\}$. Here $n$ can also be a superindex
ranging over an infinite set of discrete values (as in the case of Floquet theory which is employed in the main text,
in that case we have $n=(j\nu)$).

\mez The target is coupled to a thermal gas whose non-interacting particles live in a $d$-dimensional space $(d \ge 1)$.
An interaction between our target system and the gas is facilitated through inelastic collisions. In the regime of low number
density of the gas particles, only single particle collisions are responsible for dynamics \cite{dumcke_low_1985}.

\mez Consider now collision of a single particle (carrying a given momentum ${\bm p}$) with the target in a given state $n$.
The pertinent incoming state $| \bm{p}\,n \ra$  corresponds to energy $\frac{\bm{p}^2}{2\,m}+{\cal E}_n$. Net outcome of the
collision process is described by the scattering matrix
 \be \label{S-Appendix-B}
 \la \bm{p}'\n\,n' | \hat{S} | \bm{p}\,n \ra \; = \; \delta^d\n({\bm p}'\n-\bm{p}) \; \delta_{n'\n n}
 \; - \; 2\,\pi\,i \; \delta\m\m\left(\frac{\bm{p}'^2}{2\,m}+{\cal E}_{n'}-\frac{\bm{p}^2}{2\,m}-{\cal E}_n\right)
 \la \bm{p}'\n\,n' | \hat{T} | \bm{p}\,n \ra \;\;\; ;
 \ee
here $\la \bm{p}'\n\,n' | \hat{T} | \bm{p}\,n \ra$ are the pertinent on-shell $\hat{T}$-matrix elements.

\mez Unitary character of the quantum dynamical evolution implies unitarity of $\hat{S}$ in the scattering sector.
The purpose of this section is to show that the just mentioned unitarity of $\hat{S}$ gives rise to the thermalization
conditions referenced in the main text.

\subsection{The first unitarity property and its consequences}

 \mez The first unitarity property $\hat{S}^\dagger \hat{S} = \hat{1}$ is equivalently expressed as
 \be \label{unitarity-1}
 \int_{{\mathbb R}^d} \m {\rm d}^dp''\n \, \sum_{n''} \,
 \la \bm{p}''\n\,n'' | \hat{S} | \bm{p}\,n \ra^{\m *} \,
 \la \bm{p}''\n\,n'' | \hat{S} | \bm{p}'\n\,n' \ra \; = \; \delta^d\n({\bm p}'\n-\bm{p}) \; \delta_{n'\n n} \;\;\; .
 \ee
 Relation (\ref{unitarity-1}) holds iff the {\sl on-shell} $\hat{T}$-matrix elements satisfy
 \begin{eqnarray} \label{unitarity-1-take-1}
 & & i \, \left\{ \la \bm{p}\,n | \hat{T} | \bm{p}'\n\,n' \ra \, - \, \la \bm{p}'\n\,n' | \hat{T} | \bm{p}\,n \ra^{\m *} \right\} \; = \\
 & = & 2\,\pi \, \int_{{\mathbb R}^d} \m {\rm d}^dp'' \sum_{n''} \,
 \delta\m\m\left( \frac{\bm{p}''^2}{2\,m}+{\cal E}_{n''}-\frac{\bm{p}^2}{2\,m}-{\cal E}_n \right) \;
 \la \bm{p}''\n\,n'' | \hat{T} | \bm{p}\,n \ra^{\m *} \, \la \bm{p}''\n\,n'' | \hat{T} | \bm{p}'\n\,n' \ra \;\;\; . \nonumber
 \end{eqnarray}
 Switch into the spherical polar coordinates, such that
 \be \label{polar}
 \int_{{\mathbb R}^d} \m {\rm d}^dp''\n \; \longleftrightarrow \; \int_{0}^{\infty} \m p''^{d-1} \, {\rm d}p''
 \int_{{\mathbb S}^{d-1}} \m {\rm d}^{d-1}\Omega'' \mez .
 \ee
 Instead of (\ref{unitarity-1-take-1}) we have then
 \begin{eqnarray} \label{unitarity-1-take-2}
 & & i \, \left\{ \la \bm{p}\,n | \hat{T} | \bm{p}'\n\,n' \ra \, - \, \la \bm{p}'\n\,n' | \hat{T} | \bm{p}\,n \ra^{\m *} \right\} \; = \\
 & = & 2\,\pi\,m \; \sum_{n''} \, \Theta\Bigl({\bm p}^2 + 2\,m\,({\cal E}_n-{\cal E}_{n''}\m)\Bigr) \;
 \tilde{p}_{n''}^{\,d-2} \int_{{\mathbb S}^{d-1}} \m\m {\rm d}^{d-1}\Omega'' \;
 \la \tilde{p}_{n''}\,\bm{\Omega}''\n\,n'' | \hat{T} | \bm{p}\,n \ra^{\m *} \,
 \la \tilde{p}_{n''}\,\bm{\Omega}''\n\,n'' | \hat{T} | \bm{p}'\n\,n' \ra \;\;\; ; \nonumber
 \end{eqnarray}

 where we have used an auxiliary shorthand symbol
 \be \label{tilde-p}
 \tilde{p}_{n''} \; = \; +\,\sqrt{ {\bm p}^2 + 2\,m\,({\cal E}_n-{\cal E}_{n''}\m) } \; = \;
 +\,\sqrt{ {\bm p}'^2 + 2\,m\,({\cal E}_{n'}\n-{\cal E}_{n''}\m) } \mez .
 \ee
 Set now $(\bm{p}'\n\,n'\n)=(\bm{p}\,n)$, this yields a simpler formula
 \begin{eqnarray} \label{unitarity-1-take-3}
 & & i \, \left\{ \la \bm{p}\,n | \hat{T} | \bm{p}\,n \ra \, - \, \la \bm{p}\,n | \hat{T} | \bm{p}\,n \ra^{\m *} \right\} \; = \\
 & = & 2\,\pi\,m \; \sum_{n''} \, \Theta\Bigl({\bm p}^2 + 2\,m\,({\cal E}_n-{\cal E}_{n''}\m)\Bigr) \;
 \tilde{p}_{n''}^{\,d-2} \int_{{\mathbb S}^{d-1}} \m\m {\rm d}^{d-1}\Omega'' \;
 \Bigl| \, \la \tilde{p}_{n''}\,\bm{\Omega}''\n\,n'' | \hat{T} | \bm{p}\,n \ra \, \Bigr|^2 \;\;\; . \nonumber
 \end{eqnarray}
 Equation (\ref{unitarity-1-take-3}) shows that the {imaginary part of the diagonal element $\la \bm{p}\,n | \hat{T} | \bm{p}\,n \ra$
 is actually negative  and expressible using
 $\Bigl| \, \la \tilde{p}_{n''}\,\bm{\Omega}''\n\,n'' | \hat{T} | \bm{p}\,n \ra \, \Bigr|^2$. 

\subsection{The second unitarity property and its consequences}

\mez The second unitarity property $\hat{S} \, \hat{S}^\dagger = \hat{1}$ is equivalently expressed as
 \be \label{unitarity-2}
 \int_{{\mathbb R}^d} \m {\rm d}^dp''\n \, \sum_{n''} \,
 \la \bm{p}\,n | \hat{S} | \bm{p}''\n\,n'' \ra^{\m *} \,
 \la \bm{p}'\n\,n' | \hat{S} | \bm{p}''\n\,n'' \ra \; = \; \delta^d\n({\bm p}'\n-\bm{p}) \; \delta_{n'\n n} \;\;\; .
 \ee
 Relation (\ref{unitarity-2}) holds iff the {\sl on-shell} $\hat{T}$-matrix elements satisfy
 \begin{eqnarray} \label{unitarity-2-take-1}
 & & i \, \left\{ \la \bm{p}'\n\,n' | \hat{T} | \bm{p}\,n \ra \, - \, \la \bm{p}\,n | \hat{T} | \bm{p}'\n\,n' \ra^{\m *} \right\} \; = \\
 & = & 2\,\pi \, \int_{{\mathbb R}^d} \m {\rm d}^dp'' \sum_{n''} \,
 \delta\m\m\left( \frac{\bm{p}''^2}{2\,m}+{\cal E}_{n''}-\frac{\bm{p}^2}{2\,m}-{\cal E}_n \right) \;
 \la \bm{p}\,n | \hat{T} | \bm{p}''\n\,n'' \ra^{\m *} \, \la \bm{p}'\n\,n' | \hat{T} | \bm{p}''\n\,n'' \ra \;\;\; . \nonumber
 \end{eqnarray}
 Switch again into the spherical polar coordinates following (\ref{polar}),
 and take advantage of our shorthand symbol $\tilde{p}_{n''}=(\ref{tilde-p})$.
 Instead of (\ref{unitarity-2-take-1}) we have then
 \begin{eqnarray} \label{unitarity-2-take-2}
 & & i \, \left\{ \la \bm{p}'\n\,n' | \hat{T} | \bm{p}\,n \ra \, - \, \la \bm{p}\,n | \hat{T} | \bm{p}'\n\,n' \ra^{\m *} \right\} \; = \\
 & = & 2\,\pi\,m \; \sum_{n''} \, \Theta\Bigl({\bm p}^2 + 2\,m\,({\cal E}_n-{\cal E}_{n''}\m)\Bigr) \;
 \tilde{p}_{n''}^{\,d-2} \int_{{\mathbb S}^{d-1}} \m\m {\rm d}^{d-1}\Omega'' \;
 \la \bm{p}\,n | \hat{T} | \tilde{p}_{n''}\,\bm{\Omega}''\n\,n'' \ra^{\m *} \,
 \la \bm{p}'\n\,n' | \hat{T} | \tilde{p}_{n''}\,\bm{\Omega}''\n\,n'' \ra \;\;\; . \nonumber
 \end{eqnarray}
 Set now $(\bm{p}'\n\,n'\n)=(\bm{p}\,n)$, this yields a simpler formula
 \begin{eqnarray} \label{unitarity-2-take-3}
 & & i \, \left\{ \la \bm{p}\,n | \hat{T} | \bm{p}\,n \ra \, - \, \la \bm{p}\,n | \hat{T} | \bm{p}\,n \ra^{\m *} \right\} \; = \\
 & = & 2\,\pi\,m \; \sum_{n''} \, \Theta\Bigl({\bm p}^2 + 2\,m\,({\cal E}_n-{\cal E}_{n''}\m)\Bigr) \;
 \tilde{p}_{n''}^{\,d-2} \int_{{\mathbb S}^{d-1}} \m\m {\rm d}^{d-1}\Omega'' \;
 \Bigl| \, \la \bm{p}\,n | \hat{T} | \tilde{p}_{n''}\,\bm{\Omega}''\n\,n'' \ra \, \Bigr|^2 \;\;\; . \nonumber
 \end{eqnarray}
 Equation (\ref{unitarity-2-take-3}) shows that theimaginary part of the diagonal element $\la \bm{p}\,n | \hat{T} | \bm{p}\,n \ra$
 is actually  negative  and expressible using
 $\Bigl| \, \la \bm{p}\,n | \hat{T} | \tilde{p}_{n''}\,\bm{\Omega}''\n\,n'' \ra \, \Bigr|^2$.

\subsection{The two unitarity properties merged together}

\mez Combine now (\ref{unitarity-1-take-3}) and (\ref{unitarity-2-take-3}) as to get
 \begin{eqnarray} \label{merged}
 & & \sum_{n''} \, \Theta\Bigl({\bm p}^2 + 2\,m\,({\cal E}_n-{\cal E}_{n''}\m)\Bigr) \;
 \tilde{p}_{n''}^{\,d-2} \int_{{\mathbb S}^{d-1}} \m\m {\rm d}^{d-1}\Omega'' \;
 \Bigl| \, \la \tilde{p}_{n''}\,\bm{\Omega}''\n\,n'' | \hat{T} | \bm{p}\,n \ra \, \Bigr|^2 \; = \\
 & = & \sum_{n''} \, \Theta\Bigl({\bm p}^2 + 2\,m\,({\cal E}_n-{\cal E}_{n''}\m)\Bigr) \;
 \tilde{p}_{n''}^{\,d-2} \int_{{\mathbb S}^{d-1}} \m\m {\rm d}^{d-1}\Omega'' \;
 \Bigl| \, \la \bm{p}\,n | \hat{T} | \tilde{p}_{n''}\,\bm{\Omega}''\n\,n'' \ra \, \Bigr|^2 \;\;\; . \nonumber
 \end{eqnarray}
 Here $\Theta(\xi \ge 0)=1$ and $\Theta(\xi<0)=0$.
 The just displayed relation (\ref{merged}),
 which has been derived here just by means of the scattering theory alone,
 represents our basic lemma (or symmetry property) leading towards the desired thermalization conditions.
 Note that the only difference between the l.h.s.~and the r.h.s.~of (\ref{merged})
 consists in making an interchange $(\bm{p}\,n) \longleftrightarrow (\tilde{p}_{n''}\,\bm{\Omega}''\n\,n'')$ in the
 $\hat{T}$-matrix elements. The thermalization conditions actually come out immediately, as detailed in the next subsection.

\mez Before proceeding further, we find it useful to highlight yet another neat symmetry property which follows from (\ref{merged})
as a consequence. Namely, after integrating both sides of (\ref{merged}) over $\bm{p} \in {\mathbb R}^d$ one gets
 \begin{eqnarray} \label{merged-symmetry}
 & & \sum_{n''} \, \int_{{\mathbb R}^d} \m {\rm d}^dp \int_{{\mathbb R}^d} \m {\rm d}^dp''\n \;
 \delta\m\m\left(\frac{\bm{p}''^2}{2\,m}+{\cal E}_{n''}-\frac{\bm{p}^2}{2\,m}-{\cal E}_n\right) \;
 \Bigl| \, \la \bm{p}''\n\,n'' | \hat{T} | \bm{p}\,n \ra \, \Bigr|^2 \; = \\
 & = & \sum_{n''} \, \int_{{\mathbb R}^d} \m {\rm d}^dp \int_{{\mathbb R}^d} \m {\rm d}^dp''\n \;
 \delta\m\m\left(\frac{\bm{p}''^2}{2\,m}+{\cal E}_{n''}-\frac{\bm{p}^2}{2\,m}-{\cal E}_n\right) \;
 \Bigl| \, \la \bm{p}\,n | \hat{T} | \bm{p}''\n\,n'' \ra \, \Bigr|^2 \;\;\; . \nonumber
 \end{eqnarray}
Note again that the only difference between the l.h.s.~and the r.h.s.~of (\ref{merged-symmetry})
 consists in making an interchange $(\bm{p}\,n) \longleftrightarrow (\bm{p}''\m\,n'')$ in the
 $\hat{T}$-matrix elements.

\subsection{Thermalization conditions}

\mez Thermodynamics enters the game only at this stage.
 Take equation (\ref{merged}), multiply both sides by $m^{-1} \, e^{-\beta\bigl(\frac{\bm{p}^2}{2m}\,+\,{\cal E}_n\bigr)}$\m,
 and integrate over $\bm{p} \in {\mathbb R}^d$\n. One gets
 \begin{eqnarray} \label{thc-take-1}
 \hspace*{-1.90cm}
 & & m^{-1} \, \sum_{n''} \; \int_{{\mathbb R}^d} \m {\rm d}^dp \;\, e^{-\beta\bigl(\frac{\bm{p}^2}{2m}\,+\,{\cal E}_n\bigr)} \,
 \Theta\Bigl({\bm p}^2 + 2\,m\,({\cal E}_n-{\cal E}_{n''}\m)\Bigr) \;
 \tilde{p}_{n''}^{\,d-2} \nonumber\\ & & \int_{{\mathbb S}^{d-1}} \m\m {\rm d}^{d-1}\Omega'' \;
 \Bigl| \, \la \tilde{p}_{n''}\,\bm{\Omega}''\n\,n'' | \hat{T} | \bm{p}\,n \ra \, \Bigr|^2 \; = \\
 \hspace*{-1.90cm}
 & = & m^{-1} \, \sum_{n''} \; \int_{{\mathbb R}^d} \m {\rm d}^dp \;\, e^{-\beta\bigl(\frac{\bm{p}^2}{2m}\,+\,{\cal E}_n\bigr)} \,
 \Theta\Bigl({\bm p}^2 + 2\,m\,({\cal E}_n-{\cal E}_{n''}\m)\Bigr) \;
 \tilde{p}_{n''}^{\,d-2} \nonumber\\ & & \int_{{\mathbb S}^{d-1}} \m\m {\rm d}^{d-1}\Omega'' \;
 \Bigl| \, \la \bm{p}\,n | \hat{T} | \tilde{p}_{n''}\,\bm{\Omega}''\n\,n'' \ra \, \Bigr|^2 \;\;\; . \nonumber
 \end{eqnarray}
 As one may easily verify via direct calculation, the l.h.s.~of (\ref{thc-take-1}) equals to
 \be
 \sum_{n''} \, \int_{{\mathbb R}^d} \m {\rm d}^dp \int_{{\mathbb R}^d} \m {\rm d}^dp''\n \;
 e^{-\beta\bigl(\frac{\bm{p}^2}{2m}\,+\,{\cal E}_n\bigr)} \;
 \delta\m\m\left(\frac{\bm{p}''^2}{2\,m}+{\cal E}_{n''}-\frac{\bm{p}^2}{2\,m}-{\cal E}_n\right) \;
 \Bigl| \, \la \bm{p}''\n\,n'' | \hat{T} | \bm{p}\,n \ra \, \Bigr|^2 \mez ;
 \ee
 or, in other words, to
 \be \label{thc-take-2}
 \sum_{n''} \, a_{n''\n n} \; e^{-\beta{\cal E}_n} \;\;\; ;
 \ee
 where $a_{n''\n n}$'s are recognized as the transition rates of the main text
 (defined here up to an irrelevant multiplicative prefactor).

\mez Let us work out in a similar fashion also the r.h.s.~of (\ref{thc-take-1}). One obtains
 \begin{eqnarray}
 \hspace*{-1.00cm} & & \sum_{n''} \, \int_{{\mathbb R}^d} \m {\rm d}^dp \int_{{\mathbb R}^d} \m {\rm d}^dp''\n \;
 e^{-\beta\bigl(\frac{\bm{p}^2}{2m}\,+\,{\cal E}_n\bigr)} \;
 \delta\m\m\left(\frac{\bm{p}''^2}{2\,m}+{\cal E}_{n''}-\frac{\bm{p}^2}{2\,m}-{\cal E}_n\right) \;
 \Bigl| \, \la \bm{p}\,n | \hat{T} | \bm{p}''\n\,n'' \ra \, \Bigr|^2 \; = \nonumber\\
 \hspace*{-1.00cm} & = & \sum_{n''} \, \int_{{\mathbb R}^d} \m {\rm d}^dp \int_{{\mathbb R}^d} \m {\rm d}^dp''\n \;
 e^{-\beta\bigl(\frac{\bm{p}''^2}{2m}\,+\,{\cal E}_{n''}\n\bigr)} \;
 \delta\m\m\left(\frac{\bm{p}''^2}{2\,m}+{\cal E}_{n''}-\frac{\bm{p}^2}{2\,m}-{\cal E}_n\right) \;
 \Bigl| \, \la \bm{p}\,n | \hat{T} | \bm{p}''\n\,n'' \ra \, \Bigr|^2 \;\;\; ;
 \end{eqnarray}
 or, in other words,
 \be \label{thc-take-3}
 \sum_{n''} \, a_{n n''} \; e^{-\beta{\cal E}_{n''}} \;\;\; .
 \ee
 Since the l.h.s.~and the r.h.s.~of (\ref{thc-take-1}) are equal, we have $(\ref{thc-take-2})=(\ref{thc-take-3})$.
 Showing that, finally,
 \be \label{thc-take-4}
 \sum_{n''} \, a_{n n''} \; e^{-\beta{\cal E}_{n''}} \; = \; \sum_{n''} \, a_{n''\n n} \; e^{-\beta{\cal E}_n} \;\;\; ;
 \ee
 and these are indeed the sought thermalization conditions which we wanted to establish.
 In passing we note that the symmetry property (\ref{merged-symmetry}) is nothing else than equation (\ref{thc-take-4}) for $\beta=0$.

\section{The regime of high temperatures $\bm{(\beta \to 0)}$}

\mez The  Floquet thermalization conditions discussed above, Eq.~\ref{thc-take-4} and Eq.~(5) in the main text, enable us to draw strong conclusions
regarding the properties of the NESS of our driven system at high temperatures $(\beta \to 0)$.
Recall that the NESS is generally obtained from the Pauli NESS equation and depends on the rates $a_{j'\n j}$. In this section we explore the behavior of these rates in the regime of $\beta \to 0$.

\subsubsection{Preparations}

\mez Consider the momentum integrals
     \begin{eqnarray} \label{momentum-integrals}
        \int_{{\mathbb R}^d} \m {\rm d}^dp \int_{{\mathbb R}^d} \m {\rm d}^dp' \; e^{-\beta\frac{{\bm p}^2}{2m}} \;
        \delta\m\m\left(\frac{\bm{p}'^2}{2\,m}+E_{j'\n\nu}^{\rm QE}-\frac{\bm{p}^2}{2\,m}-E_{j0}^{\rm QE}\right) \,
        \Bigl| \la\m\n\la \bm{p}'\n\,(j'\n\nu) |\n| \widehat{T} |\n| \bm{p}\,(j0) \ra\m\n\ra \Bigr|^2
     \end{eqnarray}
in $\beta \to 0$ regime. These momentum integrals scale for $\beta \to 0$ as $\beta^{-\zeta}$, where the exponent $\zeta \ge 0$
depends upon concrete nature of the quantum system -- particle coupling. In particular, value of $\zeta=0$ refers to situations when the Floquet
$\widehat{T}$-matrix elements $| \la\m\n\la \bm{p}'\n\,(j'\n\nu) |\n| \widehat{T} |\n| \bm{p}\,(j0) \ra\m\n\ra|^2$ fall off to zero
rapidly enough at high energies, such that the entities (\ref{momentum-integrals}) converge at $\beta=0$. We shall however allow for more general
setups with $\zeta>0$, where integrals (\ref{momentum-integrals}) converge only due to presence of the Boltzmann factor
$e^{-\beta\frac{{\bm p}^2}{2m}}$ (with $\beta>0$ serving as an ultraviolet regulator).

\mez Observe now that the Floquet rates $a_{j'\n j}^{\nu}$ of equation (\ref{Floquet-rates-Appendix-A}) scale for $\beta \to 0$
as $\beta^{d/2}\,\beta^{-\zeta}$, where $\beta^{d/2}$ arises due to presence of the partition function prefactor $Z^{-1}$\m.
This motivates us to set conveniently
\be \label{a-beta-scaling}
a_{j'\n j}^{\nu} \; = \; Z^{-1}\,\beta^{-\zeta}\,\tilde{a}_{j'\n j}^{\nu} \mez ;
\ee
where $\tilde{a}_{j'\n j}^{\nu}$ is finite at $\beta=0$,
and even Taylor expandable around $\beta=0$. Such that
\be \label{tilde-a-Floquet}
   \tilde{a}_{j'\n j}^{\nu} \; = \; \tilde{a}_{j'\n j}^{\nu}\Bigr|_{\beta=0} + \;
   \beta \, \frac{{\rm d}}{{\rm d}\beta} \, \tilde{a}_{j'\n j}^{\nu}\Bigr|_{\beta=0} + \; {\cal O}(\beta^2) \;\;\; .
\ee
Analogous statements apply then of course also for the total rates
\be \label{total-rates-def}
   a_{j'\n j} \; = \; \sum_{\nu=-\infty}^{\nu=+\infty} a_{j'\n j}^\nu \mez .
\ee
Namely, one writes
$a_{j'\n j}=Z^{-1}\,\beta^{-\zeta}\,\tilde{a}_{j'\n j}$, with
\be \label{tilde-a}
   \tilde{a}_{j'\n j} \; = \; \sum_{\nu=-\infty}^{\nu=+\infty} \tilde{a}_{j'\n j}^{\nu} \; = \; \tilde{a}_{j'\n j}\Bigr|_{\beta=0} + \;
   \beta \, \frac{{\rm d}}{{\rm d}\beta} \, \tilde{a}_{j'\n j}\Bigr|_{\beta=0} + \; {\cal O}(\beta^2) \;\;\; .
\ee

\mez The Floquet thermalization conditions (5) can be expressed immediately in terms of the just
redefined rates (\ref{tilde-a-Floquet}). We have
      \be \label{tilde-thermalization-Floquet-general-take-2}
         \sum_{j'\n=1}^{N} \sum_{\nu'\n=-\infty}^{\nu'\n=+\infty} \, \tilde{a}_{jj'}^{-\nu'} \, e^{-\beta E_{j'\n\nu'}^{\rm QE}} \; - \;
         e^{-\beta E_{j0}^{\rm QE}} \, \sum_{j'\n=1}^N \sum_{\nu'\n=-\infty}^{\nu'\n=+\infty} \tilde{a}_{j'\n j}^{+\nu'\n} \; = \; 0 \;\;\; . \mez
         (1 \leq j \leq N)
      \ee
Setting $\beta=0$ in (\ref{tilde-thermalization-Floquet-general-take-2}) yields then an important property
      \be \label{tilde-thermalization-Floquet-general-take-2-beta=0}
         \sum_{j'\n=1}^{N} \, \tilde{a}_{jj'}\Bigr|_{\beta=0} \; - \;
         \sum_{j'\n=1}^{N} \, \tilde{a}_{j'\n j}\Bigr|_{\beta=0} \; = \; 0 \;\;\; . \mez (1 \leq j \leq N)
      \ee
Moreover, after extracting the ${\cal O}(\beta^1)$ order from (\ref{tilde-thermalization-Floquet-general-take-2}) one gets an intermediate outcome
\begin{eqnarray} \label{for-NESS-1-order}
   & & \sum_{j'\n=1}^{N} \frac{{\rm d}}{{\rm d}\beta} \, \tilde{a}_{jj'\n}\Bigr|_{\beta=0} \; - \;
   \sum_{j'\n=1}^{N} \sum_{\nu'\n=-\infty}^{\nu'\n=+\infty} \, \tilde{a}_{jj'}^{-\nu'}\Bigr|_{\beta=0} \,
   (E_{j'\n0}^{\rm QE}+\nu'\hbar\omega) \; = \\
   & = & \sum_{j'\n=1}^{N} \frac{{\rm d}}{{\rm d}\beta} \, \tilde{a}_{j'\n j\n}\Bigr|_{\beta=0} \; - \;
   \sum_{j'\n=1}^{N} \sum_{\nu'\n=-\infty}^{\nu'\n=+\infty} \, \tilde{a}_{j'\n j}^{(\nu'}\Bigr|_{\beta=0} \, E_{j0}^{\rm QE} \;\;\; ;
   \nonumber
\end{eqnarray}
which, after summing up over $j$, gives rise to another useful identity
\be \label{thermalization-beta-1}
   \sum_{j=1}^{N} \sum_{j'\n=1}^{N} \sum_{\nu'\n=-\infty}^{\nu'\n=+\infty} \, \tilde{a}_{jj'}^{\nu'}\Bigr|_{\beta=0}
   (\nu'\hbar\omega) \; = \; 0 \;\;\; .
\ee
Formula (\ref{thermalization-beta-1}) is displayed as equation (10) in the
main paper. In fact, (\ref{thermalization-beta-1}) is a bit more informative than (10), since it incorporates the $\beta$-scaling
expressed above in (\ref{a-beta-scaling}). 

\mez The just demonstrated general statement (\ref{thermalization-beta-1}) can be made even much stronger provided that the above discussed
integrals (\ref{momentum-integrals}) scale for $\beta \to 0$ as $\beta^{-\zeta}$ with $\zeta>0$. In that scenario, the dominant (and even
decisive) contribution to entities (\ref{momentum-integrals}) arises for $\beta \to 0$ due to contributions of high energy scattering,
where the leading order Born approximation is applicable.
Mathematically, this requires that the $\widehat{T}$-matrix does not decay too rapidly to zero with the momentum.
Otherwise, the dominant contribution would not come from the high energies. The behavior of the $\widehat{T}$-matrix at high energies
depends upon a particular problem under study. In the case of our toy model with $\delta$-type potentials, nontrivial scattering
occurs at arbitrarily high energies, and the just mentioned condition is thus fulfilled. Accordingly, instead of (\ref{momentum-integrals}) we are essentially dealing
with integrals
     \begin{eqnarray} \label{momentum-integrals-Born}
        \int_{{\mathbb R}^d} \m {\rm d}^dp \int_{{\mathbb R}^d} \m {\rm d}^dp' \; e^{-\beta\frac{{\bm p}^2}{2m}} \;
        \delta\m\m\left(\frac{\bm{p}'^2}{2\,m}-\frac{\bm{p}^2}{2\,m}\right) \,
        \Bigl| \la\m\n\la \bm{p}'\n\,(j'\n\nu) |\n| \widehat{V} |\n| \bm{p}\,(j0) \ra\m\n\ra \Bigr|^2 \;\;\; ;
     \end{eqnarray}
where $\widehat{V}=\hat{V}$ is the pertinent Floquet operator of the target -- particle coupling. In (\ref{momentum-integrals-Born}),
we have also replaced $\delta\m\left(\frac{\bm{p}'^2}{2\,m}+E_{j'\n\nu}^{\rm QE}-\frac{\bm{p}^2}{2\,m}-E_{j0}^{\rm QE}\right)$ by mere
 $\delta\m\left(\frac{\bm{p}'^2}{2\,m}-\frac{\bm{p}^2}{2\,m}\right)$, since the difference between $\frac{\bm{p}^2}{2\,m}-E_{j\nu}^{\rm QE}$
and mere $\frac{\bm{p}^2}{2\,m}$ becomes immaterial at high scattering energies. Importantly, when combining (\ref{momentum-integrals-Born})
with the Brillouin zone symmetry property

\be
   \la\m\n\la \bm{p}'\n\,(j'\n\nu) |\n| \widehat{V} |\n| \bm{p}\,(j0) \ra\m\n\ra \; = \;
   \la\m\n\la \bm{p}'\n\,(j'\n0) |\n| \widehat{V} |\n| \bm{p}\,(j,-\nu) \ra\m\n\ra \; = \;
   \la\m\n\la \bm{p}\,(j,-\nu) |\n| \widehat{V} |\n| \bm{p}'\n\,(j'\n0) \ra\m\n\ra^* \;\;\; ;
\ee
we find that
\be
   \Bigl| \, \la\m\n\la \bm{p}'\n\,(j'\n\nu) |\n| \widehat{V} |\n| \bm{p}\,(j0) \ra\m\n\ra \, \Bigr|^2 \; = \;
   \Bigl| \, \la\m\n\la \bm{p}\,(j,-\nu) |\n| \widehat{V} |\n| \bm{p}'\n\,(j'\n0) \ra\m\n\ra \, \Bigr|^2 \mez ;
\ee
and hence also
\begin{eqnarray}
   & & \int_{{\mathbb R}^d} \m {\rm d}^dp \int_{{\mathbb R}^d} \m {\rm d}^dp' \;
        \delta\m\m\left(\frac{\bm{p}'^2}{2\,m}-\frac{\bm{p}^2}{2\,m}\right) \,
        \Bigl| \la\m\n\la \bm{p}'\n\,(j'\n\nu) |\n| \widehat{V} |\n| \bm{p}\,(j0) \ra\m\n\ra \Bigr|^2 \; = \\
   & = & \int_{{\mathbb R}^d} \m {\rm d}^dp \int_{{\mathbb R}^d} \m {\rm d}^dp' \;
        \delta\m\m\left(\frac{\bm{p}'^2}{2\,m}-\frac{\bm{p}^2}{2\,m}\right) \,
        \Bigl| \, \la\m\n\la \bm{p}'\n\,(j,-\nu) |\n| \widehat{V} |\n| \bm{p}\,(j'\n0) \ra\m\n\ra \, \Bigr|^2 \mez . \nonumber
\end{eqnarray}
We thus infer that, inevitably,
\be \label{thermalization-beta-1-stronger}
   \tilde{a}_{j'\n j}^{+\nu}\Bigr|_{\beta=0} \; = \; \tilde{a}_{jj'}^{-\nu}\Bigr|_{\beta=0} \mez ;
\ee
and this leads to equation (13) of the main text.
In fact, (\ref{thermalization-beta-1-stronger}) is a bit more informative than (13), since it incorporates the $\beta$-scaling
expressed above in (\ref{a-beta-scaling}).

\mez Moreover, if extra symmetries on the driving are considered, it is possible to get yet another additional strong symmetry property
of the rates, namely,
\be \label{thermalization-beta-1-stronger-special}
   \tilde{a}_{jj'}^{\nu'}\Bigr|_{\beta=0} \; = \; \tilde{a}_{jj'}^{-\nu'}\Bigr|_{\beta=0} \;\;\; ; \mez (\zeta>0)
\ee
as in equation (14) of the main text.
In the case of our toy model with $\delta$-type potentials (see S6 below), property (\ref{thermalization-beta-1-stronger-special}) is equivalent to
      \be \label{thermalization-beta-1-stronger-delta}
         \Bigl| {\cal V}^{(j(+\nu))(j'0)} \Bigr|^2 \; = \; \Bigl| {\cal V}^{(j(-\nu))(j'0)} \Bigr|^2 \mez .
      \ee
Validity of (\ref{thermalization-beta-1-stronger-delta}) has been verified for our three-level model of subsection S6, where
\be \label{cal-V-big-def-again}
 {\cal V}^{(j'\nu')(j''\nu'')} \; = \;
 \left( \frac{1}{T} \m \int_{0}^{T} \m\m {\rm d}t' \; e^{-i(\nu'\n-\nu'')\omega t'} \,
 e^{+(j'\n-j'')\frac{i}{\hbar}\frac{\lambda}{\omega} \, \sin(\omega\,\n t'\n)} \right) \,
 {\cal V}^{(j'j'')} \mez ;
 \ee
 consonantly with (\ref{crzaux}). 
See figure \ref{fig:ratespn} for a numeric confirmation of the property (\ref{thermalization-beta-1-stronger-special}).
 
\begin{figure}[h!]
\includegraphics[scale=0.65,angle=0]{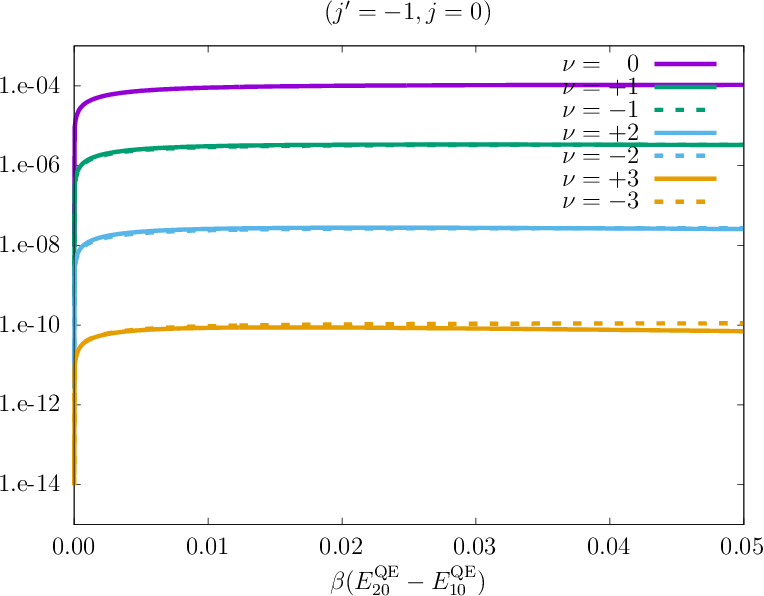} \hspace*{+0.20cm}
\includegraphics[scale=0.65,angle=0]{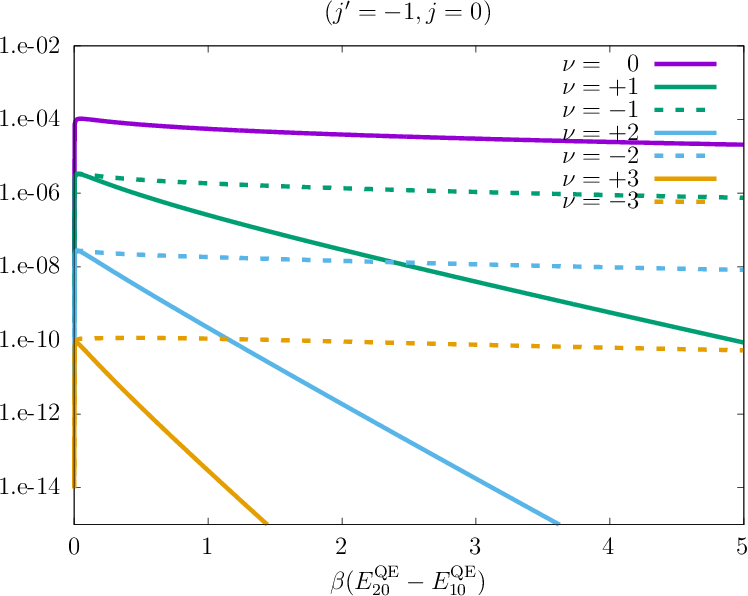}
\caption{ An example of the Floquet transition rates $a_{j'\n j}^{-\nu}$, given by equation (2) of the main text, and plotted
against the inverse temperature $\beta$. The particular rates presented here exhibit the  behavior predicted by Eq. \eqref{thermalization-beta-1-stronger-special}.
Rates for other combinations of $(j'\n, j)$ behave in a qualitatively similar manner. Parameters of the calculation are the same as those used
 for the three-level model in the main text and in S6 $(\lambda=0.5)$.
} \label{fig:ratespn}
\end{figure}

 \section{The regime of low temperatures: $\bm{(\beta \to \infty)}$}

\mez We start by recalling the definition of the rates as outlined in the manuscript:
\begin{eqnarray} \label{Floquet-ratess}
 a_{j'\n j}^{\nu} & = &
 {\cal N} \, Z^{-1} \m \int_{{\mathbb R}^d} \m {\rm d}^dp \int_{{\mathbb R}^d} \m {\rm d}^dp' \; e^{-\beta\frac{\bm{p}^2}{2\,m}} \, 
  \delta\m\m\left(\frac{\bm{p}'^2}{2\,m}+E_{j'\n\nu}^{\rm QE}-\frac{\bm{p}^2}{2\,m}-E_{j0}^{\rm QE}\right) \,
 \Bigl| \la\m\n\la \bm{p}'\n\,(j'\n\nu) |\n| \widehat{T} |\n| \bm{p}\,(j0) \ra\m\n\ra \Bigr|^2 \;\;\; .
 \end{eqnarray}
Here ${\cal N}$ stands for the gas particle density, $Z=\int_{{\mathbb R}^d} \m {\rm d}^dp \; e^{-\beta\frac{\bm{p}^2}{2\,m}}$
for the gas partition function, and $\la\m\n\la \bm{p}'\n\,(j'\n\nu) |\n| \widehat{T} |\n| \bm{p}\,(j0) \ra\m\n\ra$ are the on-shell
Floquet $\widehat{T}$-matrix elements, encoding all information about scattering of a single gas particle on the driven $N$-level system.
      For $\beta \to \infty$, the Boltzmann factor $e^{-\beta\frac{\bm{p}^2}{2\,m}}$ is significantly nonzero
      only if the kinetic energy of the incoming gas particles is very small $\left(\frac{\bm{p}^2}{2\,m} \to 0\right)$.
      Therefore, the only significant contribution to the integral (\ref{Floquet-ratess}) comes from very low energy scattering.

 \mez Let us analyze the low energy behavior of the pertinent on-shell $\widehat{T}$-matrix element
      \be \label{widehat-T-gen}
         \la\m\n\la \bm{p}'\n\,(j'\n\nu) |\n| \widehat{T} |\n| \bm{p}\,(j0) \ra\m\n\ra \; = \;
         \la\m\n\la \bm{p}'\n\,(j'\n\nu) |\n| \, \widehat{V} \, + \,
         \widehat{V} \, \frac{1}{\frac{\bm{p}^2}{2\,m}+E_{j0}^{\rm QE}-\widehat{H}+i\,0_+} \, \widehat{V} \, |\n| \bm{p}\,(j0) \ra\m\n\ra \;\;\; .
      \ee
      In the low energy regime of $\frac{\bm{p}^2}{2\,m} \to 0$, the quantity (\ref{widehat-T-gen}) reduces to
      \be \label{widehat-T-zero-T}
         \la\m\n\la \bm{p}'\n\,(j'\n\nu) |\n| \, \widehat{V} \, + \,
         \widehat{V} \, \frac{1}{E_{j0}^{\rm QE}-\widehat{H}+i\,0_+} \, \widehat{V} \, |\n| \bm{0}\,(j0) \ra\m\n\ra \;\;\; ;
      \ee
      this is a finite number independent of $\bm{p}$. Accordingly, the integral can be simplified into
      \be \label{integral-low-T-take-2}
       {\cal N} \; \int_{{\mathbb R}^d} \m {\rm d}^dp' \;
         \delta\m\m\left(\frac{\bm{p}'^2}{2\,m}+E_{j'\n\nu}^{\rm QE}-E_{j0}^{\rm QE}\right) \, \left|
         \la\m\n\la \bm{p}'\n\,(j'\n\nu) |\n| \, \widehat{V} \, + \,
         \widehat{V} \, \frac{1}{E_{j0}^{\rm QE}-\widehat{H}+i\,0_+} \, \widehat{V} \, |\n| \bm{0}\,(j0) \ra\m\n\ra \right|^2 \;\;\; .
      \ee
      If so, then our Floquet rates $a_{j'\n j}^{\nu}$ of Eq.~(\ref{Floquet-ratess}) inevitably converge to well defined zero temperature limits.
          Equation (\ref{integral-low-T-take-2}) gives zero whenever $E_{j'\n\nu}^{\rm QE}>E_{j0}^{\rm QE}$ (for channels closed at zero temperature),
      and a finite nonzero value whenever $E_{j'\n\nu}^{\rm QE}<E_{j0}^{\rm QE}$ (for channels open at zero temperature). 
      The associated total rates
      \be
         a_{j'\n j}\n(\beta\to\infty) \; = \; \sum_\nu \, a_{j'\n j}^{\nu}\n(\beta\to\infty)
      \ee
      can serve us then for determination of the low temperature NESS asymptotes $\wp_j\n(\beta\to\infty)$ 
      which we see in  Figs.~2 and 4 of the main text. We recall that  $ E_{j'\nu}^{\rm QE} = E_{j'\n 0}^{\rm QE} + \nu\hbar\omega$ where $\nu \in {\mathbb Z} $. Therefore, for every $j$ and $j'$ there exists some $\nu_0$ such that $E_{j'\nu<\nu_0}^{\rm QE}<E_{j0}^{\rm QE}$,
      meaning that $\nu<\nu_0$ are open channels contributing to $a_{j'\n j}\n(\beta\to\infty)$.
      Thus $a_{j'\n j}\n(\beta\to\infty)\neq 0$ for any $j$ and $j'$, and the pertinent NESS deviates from the ground state.
    
\section{Taylor Expansion of the steady state equation (9)}

\mez In this section we Taylor expand around $\beta=0$ the term
\begin{equation}
\sum_{j'}a_{jj'}\Bigl(\frac{\wp_{j'}^{{\rm }}}{\wp_{j}^{{\rm }}}-e^{-\beta(E_{j'0}^{QE}-E_{j0}^{QE})}\langle e^{\beta\nu\hbar\omega}\rangle_{jj'}\Bigr) \;\;\; ; \label{eq:sseq}
\end{equation}
which corresponds to l.h.s.~of equation (9) used to find the NESS.
For this purpose, we  assume that $\, a_{jj'}^{\nu}|_{\beta=0}=\, a_{jj'}^{-\nu}|_{\beta=0}$
 as in (14)
and  write the Taylor expansion of $\langle e^{\beta\nu\hbar\omega}\rangle_{jj'}$,
\begin{equation}
\langle e^{\beta\nu\hbar\omega}\rangle_{jj'}=1+\frac{\beta^{2}}{2}\frac{d^{2}}{d^{2}\beta}\langle e^{\beta\nu\hbar\omega}\rangle_{jj'} \;\;\; ;
\end{equation}
where we have already used the fact that 
\begin{equation}
\frac{d}{d^ {}\beta} \langle e^{\beta\nu\hbar\omega}\rangle_{jj'}\propto\hbar\omega\sum_{\nu=-\infty}^{\nu=+\infty}\nu\, a_{jj'}^{\nu}|_{\beta=0}=0 \;\;\; .
\end{equation}
The last equality is obtained by using Eq.~(\ref{thermalization-beta-1}).
Next, we define the Taylor expansion of the population rates,
\begin{equation}
\frac{\wp_{j'}^{{\rm }}}{\wp_{j}^{{\rm }}}=\wp_{j'j}^{{\rm \beta=0}}+\beta\wp_{j'j}^{{\rm (1)}}+\frac{\beta^{2}}{2}\wp_{j'j}^{{\rm (2)}} \;\;\; .
\end{equation}
The zero-order expansion reads as
\begin{equation}
\sum_{\nu}Z^{-1}\beta^{-\zeta}\tilde{a}_{jj'}^{\nu}|_{\beta=0}\left(\wp_{j'j}^{{\rm \beta=0}}-1\right) \;\;\; .
\end{equation}
The first order is
\begin{equation}
\beta\sum_{\nu'}Z^{-1}\beta^{-\zeta}\frac{d}{d\beta}\tilde{a}_{jj'}^{\nu'}|_{\beta=0}\left(\wp_{j'j}^{{\rm \beta=0}}-1\right)+\sum_{\nu'}Z^{-1}\beta^{-\zeta}\tilde{a}_{jj'}^{\nu'}|_{\beta=0}\beta\left(\wp_{j'j}^{{\rm (1)}}+(E_{j'0}^{QE}-E_{j0}^{QE})\right) \;\;\; .
\end{equation}
And the second order expansion comes out as
\begin{gather}
\frac{\beta^{2}}{2}\sum_{\nu}Z^{-1}\beta^{-\zeta}\frac{d^{2}}{d^{2}\beta}\tilde{a}_{jj'}^{\nu}|_{\beta=0}\left(\wp_{j'j}^{{\rm \beta=0}}-1\right)+\beta\sum_{\nu'}Z^{-1}\beta^{-\zeta}\frac{d}{d\beta}\tilde{a}_{jj'}^{\nu'}|_{\beta=0}\beta\left(\wp_{j'j}^{{\rm (1)}}+(E_{j'0}^{QE}-E_{j0}^{QE})\right)+\nonumber \\
\sum_{\nu'}Z^{-1}\beta^{-\zeta}\tilde{a}_{jj'}^{\nu'}|_{\beta=0}\frac{\beta^{2}}{2}\left(\wp_{j'j}^{{\rm (2)}}-(E_{j'0}^{QE}-E_{j0}^{QE})^{2}-\frac{d^{2}}{d^{2}\beta}\langle e^{\beta\nu\hbar\omega}\rangle_{jj'}\right) \;\;\; .
\end{gather}
From these equations one can infer that, at steady state,
\begin{equation} \label{david-criterion-supp}
\frac{\wp_{j'}^{{\rm NESS}}}{\wp_{j}^{{\rm NESS}}}=1-\beta(E_{j'0}^{QE}-E_{j0}^{QE})+\frac{\beta^{2}}{2}\left((E_{j'0}^{QE}-E_{j0}^{QE})^{2}+\frac{d^{2}}{d^{2}\beta}\langle e^{\beta\nu\hbar\omega}\rangle_{jj'}\right)+O(\beta^{3}) \;\;\; .
\end{equation}
Equation (\ref{david-criterion-supp}) leads immediately to condition (17) from the main text.

\section{Toy model and computational algorithm}

\mez In this section we define the toy model and numerical algorithm that we used to test our analytical results. We emphasize however that our analytic results are valid for any non-degenerate $N$-level system and for any periodic driving.

\subsection{Free two level system}

\mez The system Hamiltonian reads as
\be
   \hat{H}_{\rm T} \; = \; \sum_{j=1}^2 \, | j \ra_{\rm T} \; E_j \; _{\rm T}\la j | \mez ;
\ee
where subscript $_{\rm T}$ refers to "target". In our numerical calculations, we set $E_1 = -0.25$, $E_2=+0.25$.\\
Using an alternative notation, one may write equivalently
\begin{eqnarray} \label{hat-H-t-T-free-David-TLS}
 \hat{H}_{\rm T} & = & | \varphi_{2} \ra_{\rm T} \; E_2 \; _{\rm T}\la \varphi_{2} | \\
 & + & | \varphi_{1} \ra_{\rm T} \; E_1 \;
 _{\rm T}\la \varphi_{1} | \mez . \nonumber
 \end{eqnarray}

\subsection{Free three level system}

\mez The system Hamiltonian reads as
\be
   \hat{H}_{\rm T} \; = \; \sum_{j=1}^3 \, | j \ra_{\rm T} \; E_j \; _{\rm T}\la j | \mez ;
\ee
where subscript $_{\rm T}$ refers to "target". In our numerical calculations, we set $E_1 = -0.5$, $E_2=0.0$, $E_3=+0.4$.\\
Using an alternative notation, one may write equivalently
\begin{eqnarray} \label{hat-H-t-T-free-David}
 \hat{H}_{\rm T} & = & | \varphi_{3} \ra_{\rm T} \; E_3 \;
 _{\rm T}\la \varphi_{3} | \nonumber\\
 & + & | \varphi_{2} \ra_{\rm T} \; E_2 \; _{\rm T}\la \varphi_{2} | \\
 & + & | \varphi_{1} \ra_{\rm T} \; E_1 \;
 _{\rm T}\la \varphi_{1} | \mez . \nonumber
 \end{eqnarray}

\subsection{Periodically driven two level system}

We choose a simple periodic driving corresponding e.g.~to a laser driven two-level atom.
The Hamiltonian of our driven two level system takes the form
\begin{eqnarray} \label{H-S-t-TLS}
 \hat{H}_{\rm S}\n(t) & = & E_{2} \, |\varphi_2\ra_{\rm T}\n\la\varphi_2| \; + \; E_1\,|\varphi_1\ra_{\rm T}\n\la\varphi_1| \nonumber \\
 & + & \lambda \cos(\omega t) \, |\varphi_1\ra_{\rm T}\n\la\varphi_2| \; + \; \lambda \cos(\omega t) \, |\varphi_2\ra_{\rm T}\n\la\varphi_1| \;\; .
\end{eqnarray}
In our numerical calculations, the driving frequency $\omega=0.7$, and the driving strength $\lambda=0.5$.\\
The associated Floquet eigenproblem \cite{peskin1994time,lefebvre2005scattering} reads as
 \be
 \Bigl( \hat{H}_{\rm T}\n(t'\n) \, - \, i\hbar\,\partial_{t'} \Bigr) \, |\phi_{j0}(t'\n)\ra_{\rm T} \; = \;
 E_{j0}^{\rm QE} \, |\phi_{j0}(t'\n)\ra_{\rm T} \mez ; \mez j \in \{ 1,2,3 \}
 \ee
and it is easily resolved in a numerically exact manner using the standard Fourier basis set $\{e^{+i n \omega t'}\}_{n \in {\mathbb Z}}$.

 \subsection{Periodically driven three level system}

\mez We choose a simple periodic driving
 which does not invoke coupling between the eigenstates of the target.  The Hamiltonian of our driven three level system takes the form
 \begin{eqnarray} \label{hat-H-t-T-David}
 \hat{H}_{\rm T}\n(t) & = & | \varphi_{3} \ra_{\rm T} \; \Bigl( E_3 \, + \, \lambda\,\cos(\omega\,\n t) \Bigr) \;
 _{\rm T}\la \varphi_{3} | \nonumber\\
 & + & | \varphi_{2} \ra_{\rm T} \; \hspace*{+0.25cm} E_2 \hspace*{+2.20cm} \; _{\rm T}\la \varphi_{2} | \\
 & + & | \varphi_{1} \ra_{\rm T} \; \Bigl( E_1 \, - \, \lambda\,\cos(\omega\,\n t) \Bigr) \;
 _{\rm T}\la \varphi_{1} | \mez . \nonumber
 \end{eqnarray}
 In our numerical calculations, the driving frequency $\omega=1.35$, and the driving strength $\lambda=0.5$ unless stated otherwise.

\mez The associated Floquet eigenproblem \cite{peskin1994time,lefebvre2005scattering} can be solved analytically in the $t'$-domain.
 We write as usual
 \be
 \Bigl( \hat{H}_{\rm T}\n(t'\n) \, - \, i\hbar\,\partial_{t'} \Bigr) \, |\phi_{j0}(t'\n)\ra_{\rm T} \; = \;
 E_{j0}^{\rm QE} \, |\phi_{j0}(t'\n)\ra_{\rm T} \mez ;
 \ee
 where $j \in \{ 1,2,3 \}$. Let us work out separately the case of $j=3$, that is,
 \be \label{Floquet-EVP-+1}
 \Bigl( \hat{H}_{\rm T}\n(t'\n) \, - \, i\hbar\,\partial_{t'} \Bigr) \, |\phi_{30}(t'\n)\ra_{\rm T} \; = \;
 E_{30}^{\rm QE} \, |\phi_{30}(t'\n)\ra_{\rm T} \mez .
 \ee
 Explicit appearance of the Hamiltonian (\ref{hat-H-t-T-David}) enables us to employ an obvious ansatz
 \be \label{phi-+1-ansatz}
 |\phi_{30}(t'\n)\ra_{\rm T} \; = \; e^{+\frac{i}{\hbar}f_{3}\n(t')} \, | \varphi_{3} \ra_{\rm T} \mez ;
 \ee
 here the as yet unknown factor $f_{3}\n(t')$ is required to be $t'$-periodic.
 Substitution of (\ref{phi-+1-ansatz}) into (\ref{Floquet-EVP-+1}) yields the condition
 \be
 E_3 \; + \; \lambda\,\cos(\omega\,\n t'\n) \; + \; \partial_{t'} f_{3}\n(t') \; = \; E_{30}^{\rm QE} \mez .
 \ee
 Showing that (\ref{Floquet-EVP-+1}) holds iff
 \be
 f_{3}\n(t') \; = \; \Bigl( E_{30}^{\rm QE} \, - \, E_3 \Bigr) \, t' \; - \;
 \frac{\lambda}{\omega} \, \sin(\omega\,\n t'\n) \; + \; {\cal C} \mez ;
 \ee
 where ${\cal C}$ is a constant of integration.
 The already mentioned requirement of $t'$-periodicity of $f_{3}\n(t')$ implies however ${\cal C}=0$, and also
 \be
 E_{30}^{\rm QE} \; = \; E_3 \mez .
 \ee
 Then of course
 \be
 f_{3}\n(t') \; = \; - \, \frac{\lambda}{\omega} \, \sin(\omega\,\n t'\n) \mez ;
 \ee
 and, accordingly,
 \be
 |\phi_{30}(t'\n)\ra_{\rm T} \; = \; e^{-\frac{i}{\hbar}\frac{\lambda}{\omega} \, \sin(\omega\,\n t'\n)} \, | \varphi_{3} \ra_{\rm T} \mez .
 \ee
 Our $j=3$ Floquet eigenproblem (\ref{Floquet-EVP-+1}) is thus explicitly resolved, with the final outcome
 \be
 E_{30}^{\rm QE} \; = \; E_3 \mez , \mez
 |\phi_{30}(t'\n)\ra_{\rm T} \; = \; e^{-\frac{i}{\hbar}\frac{\lambda}{\omega} \, \sin(\omega\,\n t'\n)} \, | \varphi_{3} \ra_{\rm T} \mez .
 \ee
 Completely analogous reasoning gives the remaining Floquet eigensolutions, namely,
 \be
 E_{10}^{\rm QE} \; = \; E_1 \mez , \mez
 |\phi_{10}(t'\n)\ra_{\rm T} \; = \; e^{+\frac{i}{\hbar}\frac{\lambda}{\omega} \, \sin(\omega\,\n t'\n)} \, | \varphi_{1} \ra_{\rm T} \mez ;
 \ee
 and also
 \be
 E_{20}^{\rm QE} \; = \; E_2 \mez , \mez |\phi_{20}(t'\n)\ra_{\rm T} \; = \; | \varphi_{2} \ra_{\rm T} \mez .
 \ee
 For the sake of completeness and clarity, let us also comment here on the issue of Brillouin copies. One has
 \be
 | \phi_{j\nu}(t'\n)\ra_{\rm T} \; = \; e^{+ i \nu \omega t'} \, | \phi_{j0}(t'\n)\ra_{\rm T} \mez ;
 \ee
 with the associated quasi-energy eigenvalue being equal to $E_{j0}^{\rm QE}+\nu\,\hbar\omega$.

 \mez The Floquet problem corresponding to Hamiltonian (\ref{hat-H-t-T-David}) is now completely resolved.
 We recall that the $E_{j0}^{\rm QE}$'s have just been found above.
 Concerning the eigenvectors, we write jointly
 \be
 | \phi_{j\nu}(t'\n)\ra_{\rm T} \; = \; e^{+i \nu \omega t'} \,
 e^{-(j-2)\frac{i}{\hbar}\frac{\lambda}{\omega} \, \sin(\omega\,\n t'\n)} \, | \varphi_{j} \ra_{\rm T} \mez .
 \ee
 Instead of the $| \phi_{j\nu}(t'\n)\ra_{\rm T}$ notation one may take advantage of the double bracket notation
 $|\n| \phi_{j\nu} \ra\m\ra$ where the extra ket bracket refers to the $t'$ coordinate (degree of freedom) \cite{peskin1994time}.

\subsection{Our two level system coupled to thermal gas}

In our model, we assume that the thermal gas lives in one spatial dimension \cite{alicki2023violation}. We also adopt exactly the same form of system -- bath coupling as in Ref.~\cite{alicki2023violation}. The total Hamiltonian of our two level model thus takes the form
 \be
 \hat{H}\m(t) \; = \; \hat{H}_{\rm P} \; + \; \hat{H}_{\rm T}\n(t) \; + \; \hat{V} \mez ;
 \ee
 with
 \be
 \hat{H}_{\rm P} \; = \; \frac{\hat{p}^2}{2\,m} \mez ;
 \ee
 and
 \be
 \hat{V} \; = \;
 \sum_{\iota=0}^{1} \, | \chi_\iota \ra_{\rm T} \; V_\iota \, \delta(\hat{x}-q_\iota) \; _{\rm T}\n\la \chi_\iota | \mez ;
 \ee
 where
     \begin{eqnarray}
         _{\rm T}\n\la \chi_0 | \varphi_1 \ra_{\rm T} & = & +\frac{1}{\sqrt{2}} \\
         _{\rm T}\n\la \chi_1 | \varphi_1 \ra_{\rm T} & = & +\frac{1}{\sqrt{2}} \\
         _{\rm T}\n\la \chi_0 | \varphi_2 \ra_{\rm T} & = & +\frac{1}{\sqrt{2}} \\
         _{\rm T}\n\la \chi_1 | \varphi_2 \ra_{\rm T} & = & -\frac{1}{\sqrt{2}} \mez .
      \end{eqnarray}
      In our numerical calculations, we set $V_1=0.5$, $V_2=1.0$, and $q_0=q_1=0$.
      We also set ${\cal N}=1$ for the sake of maximum simplicity.

\subsection{Our three level system coupled to thermal gas}

\mez We again assume that the thermal gas lives in one spatial dimension \cite{alicki2023violation}.
We also adopt again exactly the same form of system -- bath coupling as in Ref.~\cite{alicki2023violation}.
The total Hamiltonian of our three level model thus takes the form
 \be
 \hat{H}\m(t) \; = \; \hat{H}_{\rm P} \; + \; \hat{H}_{\rm T}\n(t) \; + \; \hat{V} \mez ;
 \ee
 with
 \be
 \hat{H}_{\rm P} \; = \; \frac{\hat{p}^2}{2\,m} \mez ;
 \ee
 and
 \be
 \hat{V} \; = \;
 \sum_{\iota=0}^{2} \, | \chi_\iota \ra_{\rm T} \; V_\iota \, \delta(\hat{x}-q_\iota) \; _{\rm T}\n\la \chi_\iota | \mez ;
 \ee
 where
     \begin{eqnarray}
         \label{efs-first}
         _{\rm T}\n\la \chi_0 | \varphi_2 \ra_{\rm T} & = & \frac{1}{\sqrt{3}} \\
         _{\rm T}\n\la \chi_1 | \varphi_2 \ra_{\rm T} & = & \frac{1}{\sqrt{3}} \\
         _{\rm T}\n\la \chi_2 | \varphi_2 \ra_{\rm T} & = & \frac{1}{\sqrt{3}} \\
         _{\rm T}\n\la \chi_0 | \varphi_1 \ra_{\rm T} & = & \frac{1}{\sqrt{3}} \\
         _{\rm T}\n\la \chi_1 | \varphi_1 \ra_{\rm T} & = & \frac{e^{+i\frac{4\pi}{3}}}{\sqrt{3}} \\
         _{\rm T}\n\la \chi_2 | \varphi_1 \ra_{\rm T} & = & \frac{e^{-i\frac{4\pi}{3}}}{\sqrt{3}} \\
         _{\rm T}\n\la \chi_0 | \varphi_3 \ra_{\rm T} & = & \frac{1}{\sqrt{3}} \\
         _{\rm T}\n\la \chi_1 | \varphi_3 \ra_{\rm T} & = & \frac{e^{+i\frac{2\pi}{3}}}{\sqrt{3}} \\
         \label{efs-last}
         _{\rm T}\n\la \chi_2 | \varphi_3 \ra_{\rm T} & = & \frac{e^{-i\frac{2\pi}{3}}}{\sqrt{3}} \mez .
      \end{eqnarray}
      In our numerical calculations, we set $V_1=\frac{1.0}{\sqrt{2\,\pi}}$, $V_2=\frac{0.7}{\sqrt{2\,\pi}}$, $V_3=\frac{1.5}{\sqrt{2\,\pi}}$,  and $q_0=q_1=q_2=0$. We also set ${\cal N}=1$ for the sake of maximum simplicity.

\mez Define
     \be
        |\n| \chi_\iota \, \nu \ra\m\ra \; = \; e^{+i \nu \omega t'} \, | \chi_\iota \ra_{\rm t} \mez ;
     \ee
     and recall that
     \be
        |\n| \phi_{j\nu} \ra\m\ra \; = \; | \phi_{j\nu}(t'\n)\ra_{\rm T} \mez .
     \ee
Moment of reflection reveals then that
 \be \label{crzaux}
    \la\m\la \chi_\iota \, \nu |\n| \phi_{j\nu} \ra\m\ra \; = \;
 \left( \frac{1}{T} \m \int_{0}^{T} \m\m {\rm d}t' \; e^{-i \nu \omega t'} \, e^{+i \nu \omega t'} \,
 e^{-(j-2)\frac{i}{\hbar}\frac{\lambda}{\omega} \, \sin(\omega\,\n t'\n)} \right) \,
 _{\rm T}\n\la \chi_\iota | \varphi_{j} \ra_{\rm T} \mez .
 \ee

\subsection{Computation of the $\bm{\widehat{T}}$-matrix elements by means of the Floquet scattering theory}

\mez Our formulation given below is obtained via merging the Floquet scattering theory of
Refs.~\cite{peskin1994time,lefebvre2005scattering} together with specific features of scattering in our toy model as detailed
in Ref.~\cite{alicki2023violation}.
In what follows, we are using either the direct product basis set
 \be
 |\n| \, x \, \phi_{j\nu} \, \ra\m\ra \; = \; | x \ra_{\rm P} \, |\n| \phi_{j\nu} \, \ra\m\ra \mz ;
 \ee
 or an equivalent direct product basis set
 \be
 |\n| \, p \, \phi_{j\nu} \, \ra\m\ra \; = \; | x \ra_{\rm P} \, |\n| \phi_{j\nu} \, \ra\m\ra \mz ;
 \ee
 here subscript $_{\rm P}$ refers to "particle". Let $|\n| \, \Psi \, \ra\m\ra$ be the total scattering eigenfunction
 of the whole ensemble "particle + target".
 We set correspondingly
 \be
 \Psi_{j\nu}\n(x) \; = \; \la\m\la \, x \, \phi_{j\nu} \, |\n| \, \Psi \, \ra\m\ra \mez , \mez
 \tilde{\Psi}_{j\nu}\n(p) \; = \; \la\m\la \, p \, \phi_{j\nu} \, |\n| \, \Psi \, \ra\m\ra \mz .
 \ee

 \mez Let us develop now the Floquet scattering formalism based upon the implicit Lippmann-Schwinger equation (LSE)
 \be \label{LSE-implicit-formal}
 |\n| \, \Psi^{(pj\nu)} \, \ra\m\ra \; = \; |\n| \, p \, \phi_{j\nu} \, \ra\m\ra \; + \;
 \frac{1}{E_{pj\nu}^{\rm QE}-\widehat{H}_0+i\varepsilon} \; \widehat{V} \, |\n| \, \Psi^{(pj\nu)} \, \ra\m\ra \mz ;
 \ee
 where
 $\widehat{H}_0=\hat{H}_{\rm P} + \hat{H}_{\rm T}\n(t'\n)$, $\widehat{V}=\hat{V}$, and
 \be \label{E-QE-full-def}
 E_{pj\nu}^{\rm QE} \; = \; \frac{p^2}{2\,m} \; + \; E_{j0}^{\rm QE} \; + \; \nu\,\hbar\omega \mz .
 \ee
 Our ultimate goal is to access the $\widehat{T}$-matrix elements
 \be \label{widehat-T-matel-formal}
 \la\m\la p'\n \,\m \phi_{j'\nu'} |\n| \, \widehat{T} \, |\n| \, p \, \phi_{j\nu} \, \ra\m\ra \; = \;
 \la\m\la p'\n \,\m \phi_{j'\nu'} |\n| \, \widehat{V} \, |\n| \, \Psi^{(pj\nu)} \, \ra\m\ra \mz .
 \ee
 In the momentum representation, the implicit LSE (\ref{LSE-implicit-formal}) takes the form
 \be \label{LSE-implicit-p-representation}
 \tilde{\Psi}_{j'\nu'}^{(pj\nu)}\n(p') \; = \;
 \delta(p-p') \, \delta_{jj'} \, \delta_{\nu\nu'} \; + \;
 \frac{1}{E_{pj\nu}^{\rm QE}-E_{p'j'\nu'}^{\rm QE}+i\varepsilon} \;
 \la\m\la p'\n \,\m \phi_{j'\nu'} |\n| \, \widehat{V} \, |\n| \, \Psi^{(pj\nu)} \, \ra\m\ra \mz . \mz [\varepsilon\to+0]
 \ee
 Direct calculation based upon (\ref{widehat-T-matel-formal}) yields explicitly the associated
 $\widehat{T}$-matrix elements. One gets
 \be \label{widehat-T-matel-explicit}
 \la\m\la p'\n \,\m \phi_{j'\nu'} |\n| \, \widehat{V} \, |\n| \, \Psi^{(pj\nu)} \, \ra\m\ra \; = \;
 \sum_{j''\nu''} \, \sum_{\iota=0}^{2} \, e^{-\frac{i}{\hbar}p'q_\iota} \m
 \left( \sum_\nu \, \la\m\la \phi_{j'\nu'} \, |\n| \chi_\iota \, \nu \ra\m\ra \; V_\iota \;
 \la\m\la  \chi_\iota \, \nu |\n| \phi_{j''\nu''} \ra\m\ra \right) \Psi_{j''\nu''}^{(pj\nu)}\n(q_\iota) \mz .
 \ee
 Showing that all the information about scattering is encoded
 just in the wavefunction amplitudes $\{ \Psi_{j''\nu''}^{(pj\nu)}\n(q_\iota) \}$.\\
 To simplify our subsequent notations, we define for convenience a shorthand symbol
 \be \label{cal-V-shorthand}
 {\cal V}_\iota^{(j'\nu')(j''\nu'')} \; = \;
 \sum_\nu \, \la\m\la \phi_{j'\nu'} \, |\n| \chi_\iota \, \nu \ra\m\ra \; V_\iota \;
 \la\m\la  \chi_\iota \, \nu | \phi_{j''\nu''} \ra\m\ra \mz .
 \ee
For the case of our two-level model, ${\cal V}_\iota^{(j'\nu')(j''\nu'')}$ is computed numerically.
 For the case of our three-level model, direct calculation yields
 \be \label{cal-V-big-def}
 {\cal V}^{(j'\nu')(j''\nu'')} \; = \;
 \left( \frac{1}{T} \m \int_{0}^{T} \m\m {\rm d}t' \; e^{-i(\nu'\n-\nu'')\omega t'} \,
 e^{+(j'\n-j'')\frac{i}{\hbar}\frac{\lambda}{\omega} \, \sin(\omega\,\n t'\n)} \right) \,
 {\cal V}^{(j'j'')} \mez ;
 \ee
 where by definition
 \be
 {\cal V}^{(j'j'')} \; = \; \sum_\iota \,
 _{\rm T}\la \varphi_{j'} | \chi_\iota \ra_{\rm T} \; V_\iota \; _{\rm T}\n\la \chi_\iota | \varphi_{j''} \ra_{\rm T} \mez .
 \ee
 In the position representation, the implicit LSE (\ref{LSE-implicit-p-representation}) takes then the form
 \begin{eqnarray} \label{LSE-implicit-x-representation}
 \Psi_{j'\nu'}^{(pj\nu)}\n(x) & = &
 \frac{e^{+\frac{i}{\hbar}px}}{\sqrt{2\,\pi\,\hbar}} \, \delta_{jj'} \, \delta_{\nu\nu'} \nonumber\\
 & + & \frac{1}{\sqrt{2\,\pi\,\hbar}} \, \sum_{\iota'} \, \int_{-\infty}^{+\infty} \m {\rm d}p' \;
 \frac{e^{+\frac{i}{\hbar}p'(x-q_{\iota'})}}{E_{pj\nu}^{\rm QE}-E_{p'j'\nu'}^{\rm QE}+i\varepsilon} \;
 \sum_{j''\nu''} {\cal V}_{\iota'}^{(j'\nu')(j''\nu'')} \,
 \Psi_{j''\nu''}^{(pj\nu)}\n(q_{\iota'}) \mz .
 \end{eqnarray}
 After setting $x=q_\iota$ in (\ref{LSE-implicit-x-representation}) one arrives towards a set of coupled linear equations
 for the unknown wavefunction amplitudes $\{ \Psi_{j''\nu''}^{(pj\nu)}\n(q_\iota) \}$. Namely, we have
 \begin{eqnarray} \label{Psi-lineq-q-iota}
 \Psi_{j'\nu'}^{(pj\nu)}\n(q_\iota) & = &
 \frac{e^{+\frac{i}{\hbar}pq_\iota}}{\sqrt{2\,\pi\,\hbar}} \, \delta_{jj'} \, \delta_{\nu\nu'} \nonumber\\
 & + & \frac{1}{\sqrt{2\,\pi\,\hbar}} \, \sum_{\iota'} \, \int_{-\infty}^{+\infty} \m {\rm d}p' \;
 \frac{e^{+\frac{i}{\hbar}p'(q_{\iota}-q_{\iota'})}}{E_{pj\nu}^{\rm QE}-E_{p'j'\nu'}^{\rm QE}+i\varepsilon} \;
 \sum_{j''\nu''} {\cal V}_{\iota'}^{(j'\nu')(j''\nu'')} \,
 \Psi_{j''\nu''}^{(pj\nu)}\n(q_{\iota'}) \mz ;
 \end{eqnarray}
 this is required to hold for all $(\iota,j',\nu')$ assuming a fixed combination of input labels $(pj\nu)$.

 \mez For any fixed $(pj\nu)$, the system of linear equations (\ref{Psi-lineq-q-iota}) is treatable numerically,
 provided that ranges of all the involved Floquet $(\nu'\n,\nu''\n)$-indices are consistently truncated.
 We recall in this context that all the numerics needs to be performed just for the case of $\nu=0$, since
 \be
 \Psi_{j'\nu'}^{(pj\nu)}\n(q_\iota) \; = \; \Psi_{j'(\nu'-\nu)}^{(pj0)}\m(q_\iota) \mz ;
 \ee
 due to the symmetry with respect to Brillouin copies.
 Moreover, the only output required for the physical theory of scattering
 is actually encoded just inside the amplitudes $\{ \Psi_{j''\nu''}^{(pj0)}\n(q_\iota) \}$.
 Having in hand $\{ \Psi_{j''\nu''}^{(pj0)}\n(q_\iota) \}$, the resulting $\widehat{T}$-matrix elements of the extended theory
 are accessed by means of (\ref{widehat-T-matel-formal}) and (\ref{widehat-T-matel-explicit}). That is,
 \be \label{widehat-T-matel-final}
 \la\m\la p'\n \,\m \phi_{j'\nu'} |\n| \, \widehat{T} \, |\n| \, p \, \phi_{j0} \, \ra\m\ra \; = \;
 \sum_{\iota} \, e^{-\frac{i}{\hbar}p'q_\iota} \, \sum_{j''\nu''} \, {\cal V}_\iota^{(j'\nu')(j''\nu'')}
 \; \Psi_{j''\nu''}^{(pj0)}\n(q_\iota) \mz .
 \ee
 Equation (\ref{widehat-T-matel-final}) represents the final outcome of the scattering formalism in the extended setup
 where $q_\iota \neq 0$.

 \mez
 Substantial simplifications result in the special case of $q_\iota=0$  $(\iota=0,1\;{\rm or}\;0,1,2)$.
 Namely, instead of (\ref{Psi-lineq-q-iota}) we have a simpler set of coupled linear equations
 \begin{eqnarray} \label{Psi-lineq-q-iota-zero}
 \Psi_{j'\nu'}^{(pj\nu)} & = &
 \frac{1}{\sqrt{2\,\pi\,\hbar}} \, \delta_{jj'} \, \delta_{\nu\nu'} \nonumber\\
 & + & \frac{1}{\sqrt{2\,\pi\,\hbar}} \, \left( \, \int_{-\infty}^{+\infty} \m
 \frac{{\rm d}p'}{E_{pj\nu}^{\rm QE}-E_{p'j'\nu'}^{\rm QE}+i\varepsilon} \, \right)
 \sum_{j''\nu''} {\cal V}^{(j'\nu')(j''\nu'')} \,
 \Psi_{j''\nu''}^{(pj\nu)} \mz ;
 \end{eqnarray}
 where by definition
 \be \label{cal-V-shorthand-sum}
 {\cal V}^{(j'\nu')(j''\nu'')} \; = \; \sum_{\iota} \, {\cal V}_\iota^{(j'\nu')(j''\nu'')}
 \mz .
 \ee
 Instead of (\ref{widehat-T-matel-final}) we have then
 \be \label{widehat-T-matel-final-q-iota-zero-v1}
 \la\m\la p'\n \,\m \phi_{j'\nu'} |\n| \, \widehat{T} \, |\n| \, p \, \phi_{j0} \, \ra\m\ra \; = \;
 \sum_{j''\nu''} \, {\cal V}^{(j'\nu')(j''\nu'')}
 \; \Psi_{j''\nu''}^{(pj0)} \mz .
 \ee
 Comparison between (\ref{Psi-lineq-q-iota-zero}) and (\ref{widehat-T-matel-final-q-iota-zero-v1}) provides an alternative and even simpler formula
 \be \label{widehat-T-matel-final-q-iota-zero-v2}
 \la\m\la p'\n \,\m \phi_{j'\nu'} |\n| \, \widehat{T} \, |\n| \, p \, \phi_{j0} \, \ra\m\ra \; = \;
 \left( \, \int_{-\infty}^{+\infty} \m \frac{{\rm d}p'}{E_{pj0}^{\rm QE}-E_{p'j'\nu'}^{\rm QE}+i\varepsilon} \, \right)^{\m\m\m-1}
 \m\m\m \left( \sqrt{2\,\pi\,\hbar} \; \Psi_{j'\nu'}^{(pj0)} \, - \, \delta_{jj'} \, \delta_{0\nu'} \right)
 \mz .
 \ee
 Note also that the 1D integrals appearing in (\ref{Psi-lineq-q-iota-zero}) and (\ref{widehat-T-matel-final-q-iota-zero-v2}) are analytically solvable.
 The resulting outcome reads as follows:
 \be \label{integral-1D-analytic}
 \int_{-\infty}^{+\infty} \m \frac{{\rm d}p'}{E_{pj\nu}^{\rm QE}-E_{p'j'\nu'}^{\rm QE}+i\varepsilon} \; = \;
 -\,i\,\pi\,\sqrt{\frac{2\,m}{\frac{p^2}{2m}+E_{j0}^{\rm QE}+\nu\,\hbar\omega-E_{j'\n 0}^{\rm QE}-\nu'\n\,\hbar\omega}}
 \mz .
 \ee
 Valid regardless upon whether the channel $(j'\nu')$ is open or closed for scattering (the square root
 $\sqrt{\frac{p^2}{2m}+E_{j0}^{\rm QE}+\nu\,\hbar\omega-E_{j'\n 0}^{\rm QE}-\nu'\n\,\hbar\omega}$ is taken here to be either real positive or positive imaginary).

 \mez Equation (\ref{Psi-lineq-q-iota-zero}) can be conveniently redisplayed in a compact final appearance
 \be \label{Psi-lineq-q-iota-zero-concise}
 \sum_{j''\nu''} {\cal A}^{p,(j'\nu'),(j''\nu'')} \,
 \Psi_{j''\nu''}^{(pj\nu)} \; = \; \delta_{jj'} \, \delta_{\nu\nu'} \mz ;
 \ee
 with
 \be \label{cal-A-def}
 {\cal A}^{p,(j'\nu'),(j''\nu'')} \; = \;
 \sqrt{2\,\pi\,\hbar} \; \delta_{j'j''} \, \delta_{\nu'\nu''} \; + \;
 i\,\pi\,\sqrt{\frac{2\,m}{\frac{p^2}{2m}+E_{j0}^{\rm QE}+\nu\,\hbar\omega-E_{j'\n 0}^{\rm QE}-\nu'\n\,\hbar\omega}} \;\,
 {\cal V}^{(j'\nu')(j''\nu'')} \mz .
 \ee
 Formulas (\ref{Psi-lineq-q-iota-zero-concise})-(\ref{cal-A-def}) and (\ref{widehat-T-matel-final-q-iota-zero-v2}) are well suited for numerical implementation.

\section{Derivation of the Floquet transition rates (\ref{Floquet-rates-Appendix-A}) from quantum scattering theory}

\mez The purpose of the present section S7 is to provide a heuristic derivation of the Floquet transition rates (\ref{Floquet-rates-Appendix-A}).
A more rigorous treatment (leading to the same formula (\ref{Floquet-rates-Appendix-A})) would follow Ref.~\cite{dumcke_low_1985}. For the sake
of maximum simplicity, we shall restrict ourselves here to the case of one spatial dimension $(d=1)$.

\subsection{Preliminaries}

\mez We consider periodic driving of frequency $\omega$ and period $T = \frac{2\,\pi}{\omega}$,
and employ complete orthonormal system of zero Brillouin zone Floquet states $| \phi_{j0}\n(t) \ra$
with quasi-energies $E_{j0}^{\rm QE}$ (the used notations are the same as above in S6,
only the subscript $_{\rm T}$ is dropped for the sake of notational compactness).
      The reduced density operator is then represented in the Floquet basis as
      \be \label{varrho-S-t}
         \hat{\varrho}_{\rm S}^t \; = \; \sum_{jj'} | \phi_{j0}\n(t) \ra \, \varrho_{jj'}^t\n \, \la \phi_{j'0}\n(t) | \mez .
      \ee
In the presence of driving (but in the absence of coupling to the bath), $\hat{\varrho}_{\rm S}^t$ evolves freely. Such that
      \be \label{varrho-S-t-free}
         \hat{\varrho}_{\rm S}^t \; = \;
         \sum_{jj'} | \phi_{j0}\n(t) \ra \, \varrho_{jj'}^0\n \; e^{-\frac{i}{\hbar}(E_{j0}^{\rm QE}-E_{j'0}^{\rm QE}\n)t} \, \la \phi_{j'0}\n(t) |
         \mez . \mez [\,{\rm free}\,]
      \ee

\mez Let us turn our attention now to the bath of non-interacting particles. Each particle is in the thermal state. Such that
      \be
         \hat{\varrho}_{\rm P} \; = \; \int_{-\infty}^{+\infty} \m {\rm d}p \; | p \ra \, \frac{e^{-\beta E_p}}{\mathcal{Z}} \, \la p | \mez ;
      \ee
      here $|p\ra$ stands for a single particle momentum eigenstate, and
      \be \label{Z-def}
         \mathcal{Z} \; = \; \delta(p=0) \, \int_{-\infty}^{+\infty} \m {\rm d}p \; e^{-\beta E_p} \mez .
      \ee
      Symbol $\delta(p)$ represents the delta function of momentum.\\
      {\it More rigorous formulation would include:\\ {\it i)} Introducing the full thermal state of the many-particle bosonic system.\\
      {\it ii)} Regularization of $\delta(p=0)$.\\
      Implementation: Gas enclosed in a finite (yet arbitrarily large) volume $V$\m\n.
      Such a more rigorous (and more technical) formulation would lead towards the same final result
      (\ref{Floquet-rates-Appendix-A}) for the Floquet transition rates.}\\
      The gas is characterized by its particle number density ${\cal N}$\m.
      Meaning that, on the average, we have ${\cal N}\,V$ gas particles inside a volume $V$\m.
      (In $d=1$ the volume $V$ corresponds actually to a length segment $L$.)

\subsection{Exploring the effect of a single collision}

      \mez Let us explore an effect of a single collision (of a single scattering event).
      Let $t$ be a given time instant just before a single stochastic collision.
      The system resides in the quantum state $\hat{\varrho}_{\rm S}^t=(\ref{varrho-S-t})$.
      The total (particle + system) density operator is built up
      using an obvious direct product prescription
      \be
         \hat{\rho}^t \; = \; \int_{-\infty}^{+\infty} \m {\rm d}p \sum_{jj'} \; | p j \ra \, \frac{e^{-\beta E_p}}{\mathcal{Z}} \, \varrho_{jj'}^t\n \, \la p j'\n | \mez .
      \ee
      Here of course $| p\,j \ra = | p \ra_{\rm particle}\,| \phi_{j0} \ra_{\rm system}$.
      During a time interval $(t,t+{\rm d}t)$, a gas particle of energy $E_p$ collides with our system with the probability
      \be \label{d-wp-d-t}
         {\rm d}P_{E_p} \; = \; {\cal N} \; \frac{\sqrt{2\,m\,E_p}}{m} \; {\rm d}t \mez .
      \ee
      Scattering theory says that such a collision changes the vector $| p j \ra$ into
      \begin{eqnarray} \label{hat-S-action}
         \hat{S}\,|p j \ra & = & | p j \ra \\
         & - & 2\,\pi\,i \int_{-\infty}^{+\infty} \m {\rm d}p'' \sum_{j''} \sum_{\nu''} \,
         \delta\n\Bigl(E_{p''}+E_{j''0}^{\rm QE}+\nu''\n\hbar\omega-E_p-E_{j0}^{\rm QE}\Bigr) \; e^{+i \nu''\n \omega t} \,
         \la\m\n\la \, p''\n \, \nu''\n \, j''\n \, |\n| \, \widehat{T} |\n| \,p\,0\,j\, \ra\m\n\ra \; | p'' j'' \ra \mz . \nonumber
      \end{eqnarray}
      Putting things together, at the time instant $t+{\rm d}t$, our (particle + system) density operator must be updated into
      \be
         \hat{\rho}^{t+{\rm d}t} \; = \; \hat{\rho}_{(1)}^{t+{\rm d}t} \; + \; \hat{\rho}_{(2)}^{t+{\rm d}t} \mez .
      \ee
      Here $\hat{\rho}_{(1)}^{t+{\rm d}t}$ corresponds to the situation when no collision took place.
      Thereby $\hat{\rho}_{(1)}^{t+{\rm d}t}$ evolves just freely, such that
      \be \label{hat-rho-(1)-def}
         \hat{\rho}_{(1)}^{t+{\rm d}t} \; = \; \int_{-\infty}^{+\infty} \m {\rm d}p \sum_{jj'} \; | p j \ra \, \Bigl( 1 - {\rm d}P_{E_p} \Bigr) \,
         \frac{e^{-\beta E_p}}{\mathcal{Z}} \, \varrho_{jj'}^t\n\,e^{-\frac{i}{\hbar}(E_{j0}^{\rm QE}-E_{j'0}^{\rm QE}\n){\rm d}t} \, \la p j'\n | \mez ;
      \ee
      {\sl cf.}~equation (\ref{varrho-S-t-free}) above.
      On the other hand, $\hat{\rho}_{(2)}^{t+{\rm d}t}$ corresponds to the situation when a single collision took place.
      Accordingly, one has
      \be \label{hat-rho-(2)-def}
         \hat{\rho}_{(2)}^{t+{\rm d}t} \; = \; \int_{-\infty}^{+\infty} \m {\rm d}p \sum_{jj'} \; \hat{S}\,| p j \ra \, {\rm d}P_{E_p} \,
         \frac{e^{-\beta E_p}}{\mathcal{Z}} \, \varrho_{jj'}^t\n \, \la p j'\n |\,\hat{S}^\dagger \mez .
      \ee

\subsection{Technical elaborations \#1}

      \mez Let us express $\hat{\rho}_{(1,2)}^{t+{\rm d}t}$ more explicitly. We have
      \begin{eqnarray} \label{hat-rho-(1)-explicit}
         \hat{\rho}_{(1)}^{t+{\rm d}t} & = & \hat{\rho}^t \\
         & - & \int_{-\infty}^{+\infty} \m {\rm d}p \sum_{jj'} \; | p j \ra \,
         \frac{e^{-\beta E_p}}{\mathcal{Z}} \, \varrho_{jj'}^t\n\,\frac{i}{\hbar}\,(E_{j0}^{\rm QE}-E_{j'0}^{\rm QE}\n)\,{\rm d}t \, \la p j'\n | \nonumber\\
         & - & \int_{-\infty}^{+\infty} \m {\rm d}p \sum_{jj'} \; | p j \ra \, {\rm d}P_{E_p} \,
         \frac{e^{-\beta E_p}}{\mathcal{Z}} \, \varrho_{jj'}^t\n \, \la p j'\n | \mez . \nonumber
      \end{eqnarray}
      In addition, after combining (\ref{hat-rho-(2)-def}) with (\ref{hat-S-action}), one gets a lengthy intermediate outcome
      \begin{eqnarray} \label{hat-rho-(2)-explicit}
         \hspace*{-1.50cm} & & \hat{\rho}_{(2)}^{t+{\rm d}t} \; = \\
         \hspace*{-1.50cm} & = & \int_{-\infty}^{+\infty} \m {\rm d}p \sum_{jj'} \; | p j \ra \, {\rm d}P_{E_p} \,
         \frac{e^{-\beta E_p}}{\mathcal{Z}} \, \varrho_{jj'}^t\n \, \la p j'\n | \nonumber\\
         \hspace*{-1.50cm} & - & \int_{-\infty}^{+\infty} \m {\rm d}p \sum_{jj'} \; 2\,\pi\,i \int_{-\infty}^{+\infty} \m {\rm d}p'' \sum_{j''} \sum_{\nu''} \,
         \delta\Bigl(E_{p''}+E_{j''0}^{\rm QE}+\nu''\n\hbar\omega-E_p-E_{j0}^{\rm QE}\Bigr) \nonumber\\ \hspace*{-1.50cm} & & e^{+i \nu''\n \omega t} \;
         \la\m\n\la p''\n \, \nu''\n \, j''\n \, |\n| \widehat{T} |\n| p \, 0 \, j \ra\m\n\ra \; | p'' j'' \ra \, {\rm d}P_{E_p} \,
         \frac{e^{-\beta E_p}}{\mathcal{Z}} \, \varrho_{jj'}^t\n \, \la p j'\n | \nonumber\\
         \hspace*{-1.50cm} & + & \int_{-\infty}^{+\infty} \m {\rm d}p \sum_{jj'} \; | p j \ra \, {\rm d}P_{E_p} \,
         \frac{e^{-\beta E_p}}{\mathcal{Z}} \, \varrho_{jj'}^t\n \; 2\,\pi\,i \int_{-\infty}^{+\infty} \m {\rm d}p'' \sum_{j''} \sum_{\nu''} \,
         \delta\Bigl(E_{p''}+E_{j''0}^{\rm QE}+\nu''\n\hbar\omega-E_p-E_{j'0}^{\rm QE}\Bigr) \nonumber\\
         \hspace*{-1.50cm} & & e^{-i \nu''\n \omega t} \; \la\m\n\la p''\n \, \nu''\n \, j''\n \, |\n| \widehat{T} |\n| p \, 0 \, j'\n \ra\m\n\ra^* \; \la p'' j'' | \nonumber\\
         \hspace*{-1.50cm} & - & \int_{-\infty}^{+\infty} \m {\rm d}p \sum_{jj'} \; 2\,\pi\,i \int_{-\infty}^{+\infty} \m {\rm d}p'' \sum_{j''} \sum_{\nu''} \,
         \delta\Bigl(E_{p''}+E_{j''0}^{\rm QE}+\nu''\n\hbar\omega-E_p-E_{j0}^{\rm QE}\Bigr) \; e^{+i \nu''\n \omega t} \,
         \la\m\n\la p''\n \, \nu''\n \, j''\n \, |\n| \widehat{T} |\n| p \, 0 \, j \ra\m\n\ra \; | p'' j'' \ra \, {\rm d}P_{E_p} \nonumber\\
         \hspace*{-1.50cm} & & \frac{e^{-\beta E_p}}{\mathcal{Z}} \, \varrho_{jj'}^t\n \, 2\,\pi\,i \int_{-\infty}^{+\infty} \m {\rm d}p''' \sum_{j'''} \sum_{\nu'''} \,
         \delta\Bigl(E_{p'''}+E_{j'''0}^{\rm QE}+\nu'''\n\hbar\omega-E_p-E_{j'0}^{\rm QE}\Bigr) \; e^{-i \nu'''\n \omega t} \,
         \la\m\n\la p'''\n \, \nu'''\n \, j'''\n \, |\n| \widehat{T} |\n| p \, 0 \, j'\n \ra\m\n\ra^* \; \la p''' j''' | \mez . \nonumber
      \end{eqnarray}
      The last line of (\ref{hat-rho-(1)-explicit}) is exactly overcompensated by the first line of (\ref{hat-rho-(2)-explicit}).
      We can thus write more concisely and more insightfully
      \begin{eqnarray} \label{derivative-hat-rho}
         \hspace*{-1.50cm} & & \frac{\rm d}{{\rm d}t} \, \hat{\rho}^t \; = \\ \hspace*{-1.50cm} & = & \int_{-\infty}^{+\infty} \m {\rm d}p \sum_{jj'} \; | p j \ra \,
         \frac{e^{-\beta E_p}}{\mathcal{Z}} \, \varrho_{jj'}^t\n\m\left(-\frac{i}{\hbar}\right)\m(E_{j0}^{\rm QE}-E_{j'0}^{\rm QE}\n) \, \la p j'\n | \nonumber\\
         \hspace*{-1.50cm} & - & \int_{-\infty}^{+\infty} \m {\rm d}p \sum_{jj'} \; 2\,\pi\,i \int_{-\infty}^{+\infty} \m {\rm d}p'' \sum_{j''} \sum_{\nu''} \,
         \delta\Bigl(E_{p''}+E_{j''0}^{\rm QE}+\nu''\n\hbar\omega-E_p-E_{j0}^{\rm QE}\Bigr) \nonumber\\ \hspace*{-1.50cm} & & e^{+i \nu''\n \omega t} \;
         \la\m\n\la p''\n \, \nu''\n \, j''\n \, |\n| \widehat{T} |\n| p \, 0 \, j \ra\m\n\ra \; | p'' j'' \ra \, {\rm d}P_{E_p} \,
         \frac{e^{-\beta E_p}}{\mathcal{Z}} \, \varrho_{jj'}^t\n \, \la p j'\n | \nonumber\\
         \hspace*{-1.50cm} & + & \int_{-\infty}^{+\infty} \m {\rm d}p \sum_{jj'} \; | p j \ra \, {\rm d}P_{E_p} \,
         \frac{e^{-\beta E_p}}{\mathcal{Z}} \, \varrho_{jj'}^t\n \; 2\,\pi\,i \int_{-\infty}^{+\infty} \m {\rm d}p'' \sum_{j''} \sum_{\nu''} \,
         \delta\Bigl(E_{p''}+E_{j''0}^{\rm QE}+\nu''\n\hbar\omega-E_p-E_{j'0}^{\rm QE}\Bigr) \nonumber\\
         \hspace*{-1.50cm} & & e^{-i \nu''\n \omega t} \; \la\m\n\la p''\n \, \nu''\n \, j''\n \, |\n| \widehat{T} |\n| p \, 0 \, j'\n \ra\m\n\ra^* \; \la p'' j'' | \nonumber\\
         \hspace*{-1.50cm} & - & \int_{-\infty}^{+\infty} \m {\rm d}p \sum_{jj'} \; 2\,\pi\,i \int_{-\infty}^{+\infty} \m {\rm d}p'' \sum_{j''} \sum_{\nu''} \,
         \delta\Bigl(E_{p''}+E_{j''0}^{\rm QE}+\nu''\n\hbar\omega-E_p-E_{j0}^{\rm QE}\Bigr) \; e^{+i \nu''\n \omega t} \,
         \la\m\n\la p''\n \, \nu''\n \, j''\n \, |\n| \widehat{T} |\n| p \, 0 \, j \ra\m\n\ra \; | p'' j'' \ra \, {\rm d}P_{E_p} \nonumber\\
         \hspace*{-1.50cm} & & \frac{e^{-\beta E_p}}{\mathcal{Z}} \, \varrho_{jj'}^t\n \, 2\,\pi\,i \int_{-\infty}^{+\infty} \m {\rm d}p''' \sum_{j'''} \sum_{\nu'''} \,
         \delta\Bigl(E_{p'''}+E_{j'''0}^{\rm QE}+\nu'''\n\hbar\omega-E_p-E_{j'0}^{\rm QE}\Bigr) \; e^{-i \nu'''\n \omega t} \,
         \la\m\n\la p'''\n \, \nu'''\n \, j'''\n \, |\n| \widehat{T} |\n| p \, 0 \, j'\n \ra\m\n\ra^* \; \la p''' j''' | \mez . \nonumber
      \end{eqnarray}
      The just derived formula (\ref{derivative-hat-rho}) represents an important intermediate step
      in our effort focused on deriving the master equation for $\hat{\varrho}_{\rm S}^t$.

\subsection{Technical elaborations \#2}

      \mez Recall (\ref{d-wp-d-t}), i.e.,
      \be
         \frac{{\rm d}P_{E_p}}{{\rm d}t} \; = \; {\cal N} \; \frac{\sqrt{2\,m\,E_p}}{m} \mez ;
      \ee
      and substitute into (\ref{derivative-hat-rho}). Let us then simplify the third line of (\ref{derivative-hat-rho}) via integrating over $p$,
      noting that $\frac{\sqrt{2\,m\,E_p}}{m}\,{\rm d}p={\rm d}E_p$. One gets
      \begin{eqnarray} \label{piece-second}
         \hspace*{-1.50cm} & - & 2\,\pi\,i \; {\cal N} \, \sum_{j j'\n j''} \sum_{\nu''} \sum_{\eta=\pm 1} \,
         \int_{-\infty}^{+\infty} \m {\rm d}p'' \; \Theta(E_{p''}+E^{\rm QE}_{j''0}+\nu''\n\hbar\omega-E^{\rm QE}_{j0}) \\
         \hspace*{-1.50cm} & & | p'' j'' \ra \, \Biggl( e^{+i \nu''\n \omega t} \,
         \la\m\n\la p''\n \, \nu''\n \, j''\n \, |\n| \widehat{T} |\n| (\eta\,p) \, 0 \, j \ra\m\n\ra
         \; \frac{e^{-\beta(E_{p''}+E^{\rm QE}_{j''0}+\nu''\n\hbar\omega-E^{\rm QE}_{j0})}}{\mathcal{Z}} \, \varrho_{jj'}^t\n \Biggr) \,
         \la (\eta\,p) j'\n | \mz ;
      \end{eqnarray}
      here $p=\sqrt{2\,m\,(E_{p''}+E^{\rm QE}_{j''0}+\nu''\n\hbar\omega-E^{\rm QE}_{j0})}$.
      Similarly, let us simplify the fourth line of (\ref{derivative-hat-rho}) via integrating over $p$, one gets
      \begin{eqnarray} \label{piece-third}
         \hspace*{-1.50cm}
         & + & 2\,\pi\,i \; {\cal N} \, \sum_{jj'\n j''} \sum_{\nu''} \sum_{\eta=\pm1} \, \int_{-\infty}^{+\infty} \m {\rm d}p'' \;
         \Theta(E_{p''}+E^{\rm QE}_{j''0}+\nu''\n\hbar\omega-E^{\rm QE}_{j'0}\n) \\
         \hspace*{-1.50cm} & & | (\eta\, p) j \ra \, \Biggl( \frac{e^{-\beta(E_{p''}+E^{\rm QE}_{j''0}+\nu''\n\hbar\omega-E^{\rm QE}_{j'0}\n)}}{\mathcal{Z}} \, \varrho_{jj'}^t\n \;
         e^{+i \nu''\n \omega t} \, \la\m\n\la p''\n \, \nu''\n \, j''\n \, |\n| \widehat{T} |\n| (\eta\, p) \, 0 \, j'\n \ra\m\n\ra^{\m *} \Biggr) \, \la p'' j'' | \mz ;
      \end{eqnarray}
      here $p=\sqrt{2\,m\,(E_{p''}+E^{\rm QE}_{j''0}+\nu''\n\hbar\omega-E^{\rm QE}_{j'0}\n)}$.
      Similarly, let us simplify also the last two lines of (\ref{derivative-hat-rho}) via integrating over $p$, one gets
      \begin{eqnarray} \label{piece-fourth}
         \hspace*{-1.50cm} & + & 4\,\pi^2 \, {\cal N} \, \sum_{jj'\n j''\n j'''} \sum_{\nu''\nu'''} \sum_{\eta=\pm1} \, \int_{-\infty}^{+\infty} \m {\rm d}p'' \;
         \Theta(E_{p''}+E^{\rm QE}_{j''0}+\nu''\n\hbar\omega-E^{\rm QE}_{j0}) \; \int_{-\infty}^{+\infty} \m {\rm d}p''' \; | p'' j'' \ra \, \la p''' j''' | \;
         \varrho_{jj'}^t\n \nonumber\\
         \hspace*{-1.50cm} & & \left\{ \; \frac{e^{-\beta(E_{p''}+E^{\rm QE}_{j''0}+\nu''\n\hbar\omega-E^{\rm QE}_{j0})}}{\mathcal{Z}} \; \delta\m\Bigl(E_{p'''}+E^{\rm QE}_{j'''0}+\nu'''\n\hbar\omega-E_{p''}-E^{\rm QE}_{j''0}-\nu''\n\hbar\omega+E^{\rm QE}_{j0}-E^{\rm QE}_{j'0}\Bigr) \right. \\
         \hspace*{-1.50cm} & & \mez e^{+i \nu''\n \omega t} \, \la\m\n\la p''\n \, \nu''\n \, j''\n \, |\n| \widehat{T} |\n| (\eta\, p) \, 0 \, j \ra\m\n\ra \;
         e^{-i \nu'''\n \omega t} \, \la\m\n\la p'''\n \, \nu'''\n \, j'''\n \, |\n| \widehat{T} |\n| (\eta\, p) \, 0 \, j'\n \ra\m\n\ra^* \; \Biggr\} \mez ; \nonumber
      \end{eqnarray}
      here $p=\sqrt{2\,m\,(E_{p''}+E^{\rm QE}_{j''0}+\nu''\n\hbar\omega-E^{\rm QE}_{j0})}$.

\mez       Putting all the ingredients together, instead of (\ref{derivative-hat-rho}) we have now
      \begin{eqnarray} \label{derivative-hat-rho-take-2}
         \hspace*{-1.50cm} & & \frac{\rm d}{{\rm d}t} \, \hat{\rho}^t \; = \\ \hspace*{-1.50cm} & = & \int_{-\infty}^{+\infty} \m {\rm d}p \sum_{jj'} \; | p j \ra \,
         \frac{e^{-\beta E_p}}{\mathcal{Z}} \, \varrho_{jj'}^t\n\m\left(-\frac{i}{\hbar}\right)\m(E^{\rm QE}_{j0}-E^{\rm QE}_{j'0}\n) \, \la p j'\n | \nonumber\\
         \hspace*{-1.50cm} & - & 2\,\pi\,i \; {\cal N} \, \sum_{j j'\n j''} \sum_{\nu''} \sum_{\eta=\pm 1} \,
         \int_{-\infty}^{+\infty} \m {\rm d}p'' \; \Theta(E_{p''}+E^{\rm QE}_{j''0}+\nu''\n\hbar\omega-E^{\rm QE}_{j0}) \nonumber\\
         \hspace*{-1.50cm} & & | p'' j'' \ra \, \Biggl( e^{+i \nu''\n \omega t} \,
         \la\m\n\la p''\n \, \nu''\n \, j''\n \, |\n| \widehat{T} |\n| (\eta\,p) \, 0 \, j \ra\m\n\ra
         \; \frac{e^{-\beta(E_{p''}+E^{\rm QE}_{j''0}+\nu''\n\hbar\omega-E^{\rm QE}_{j0})}}{\mathcal{Z}} \, \varrho_{jj'}^t\n \Biggr) \,
         \la (\eta\,p) j'\n | \nonumber\\
         \hspace*{-1.50cm}
         & + & 2\,\pi\,i \; {\cal N} \, \sum_{jj'\n j''} \sum_{\nu''} \sum_{\eta=\pm1} \, \int_{-\infty}^{+\infty} \m {\rm d}p'' \;
         \Theta(E_{p''}+E^{\rm QE}_{j''0}+\nu''\n\hbar\omega-E^{\rm QE}_{j'0}\n) \nonumber\\
         \hspace*{-1.50cm} & & | (\eta\, p) j \ra \, \Biggl( \frac{e^{-\beta(E_{p''}+E^{\rm QE}_{j''0}+\nu''\n\hbar\omega-E^{\rm QE}_{j'0}\n)}}{\mathcal{Z}} \, \varrho_{jj'}^t\n \;
         e^{+i \nu''\n \omega t} \, \la\m\n\la p''\n \, \nu''\n \, j''\n \, |\n| \widehat{T} |\n| (\eta\, p) \, 0 \, j'\n \ra\m\n\ra^{\m *} \Biggr) \, \la p'' j'' |
         \nonumber\\
         \hspace*{-1.50cm} & + & 4\,\pi^2 \, {\cal N} \, \sum_{jj'\n j''\n j'''} \sum_{\nu''\nu'''} \sum_{\eta=\pm1} \, \int_{-\infty}^{+\infty} \m {\rm d}p'' \;
         \Theta(E_{p''}+E^{\rm QE}_{j''0}+\nu''\n\hbar\omega-E^{\rm QE}_{j0}) \; \int_{-\infty}^{+\infty} \m {\rm d}p''' \; | p'' j'' \ra \, \la p''' j''' | \;
         \varrho_{jj'}^t\n \nonumber\\
         \hspace*{-1.50cm} & & \left\{ \; \frac{e^{-\beta(E_{p''}+E^{\rm QE}_{j''0}+\nu''\n\hbar\omega-E^{\rm QE}_{j0})}}{\mathcal{Z}} \; \delta\m\Bigl(E_{p'''}+E^{\rm QE}_{j'''0}+\nu'''\n\hbar\omega-E_{p''}-E^{\rm QE}_{j''0}-\nu''\n\hbar\omega+E^{\rm QE}_{j0}-E^{\rm QE}_{j'0}\Bigr) \right. \nonumber\\
         \hspace*{-1.50cm} & & \mez e^{+i \nu''\n \omega t} \, \la\m\n\la p''\n \, \nu''\n \, j''\n \, |\n| \widehat{T} |\n| (\eta\, p) \, 0 \, j \ra\m\n\ra \;
         e^{-i \nu'''\n \omega t} \, \la\m\n\la p'''\n \, \nu'''\n \, j'''\n \, |\n| \widehat{T} |\n| (\eta\, p) \, 0 \, j'\n \ra\m\n\ra^* \; \Biggr\} \mez . \nonumber
      \end{eqnarray}
      The just derived formula (\ref{derivative-hat-rho-take-2}) represents an important intermediate step
      in our effort focused on deriving the master equation for $\hat{\varrho}_{\rm S}^t$.

\subsection{Tracing out the degrees of freedom of the gas particle}

\mez In order to finalize our treatment of the effect of a single stochastic collision,
      we need to take (\ref{derivative-hat-rho-take-2}) and trace out over the particle degrees of freedom.
      Such a step is legitimate, since the particle fades away and never returns,
      and our system becomes free again (while being still driven)
      until the next collision with another particle.
      So we wish to get the reduced density operator $\hat{\varrho}_{\rm S}^{t+{\rm d}t}$\m,
      which will then later on participate in collisions with the other incoming particles.
      Tracing over the particle degrees of freedom will eliminate the singular $\mathcal{Z}$ factor (\ref{Z-def}).
      Performing the trace in the second line of (\ref{derivative-hat-rho-take-2}), one gets
      \begin{eqnarray} \label{piece-first-traced}
         + \; \sum_{j \neq j'} \; | j \ra \left(-\frac{i}{\hbar}\right)\m(E^{\rm QE}_{j0}-E^{\rm QE}_{j'0}\n) \, \varrho_{jj'}^t\n \, \la j'\n | \mez .
      \end{eqnarray}

\mez Hereafter we shall assume strictly nondegenerate Floquet spectrum
      (degeneracies modulo $\hbar\omega$ are excluded from our considerations as well, note that the same assumption has been mentioned
      also in the main paper). This is an important property enabling us to simplify our workout of the required traces.
      Performing the trace in the third line of (\ref{derivative-hat-rho-take-2}), one gets
      \begin{eqnarray} \label{piece-second-traced}
         - \, 2\,\pi\,i \; {\cal N} \, \sum_{j j'\n} \; | j \ra \left( \frac{\int_{-\infty}^{+\infty} \m {\rm d}p \;\,
         e^{-\beta E_p} \; \la\m\n\la p \, 0 \, j |\n| \widehat{T} |\n| p \, 0 \, j \ra\m\n\ra}{\int_{-\infty}^{+\infty} \m {\rm d}p \; e^{-\beta E_p}} \right) \varrho_{jj'}^t\n \, \la j'\n | \mez .
      \end{eqnarray}
      Performing the trace in the fourth line of (\ref{derivative-hat-rho-take-2}), one gets
      \begin{eqnarray} \label{piece-third-traced}
         + \, 2\,\pi\,i \; {\cal N} \, \sum_{j j'\n} \; | j \ra \left( \frac{\int_{-\infty}^{+\infty} \m {\rm d}p \;\,
         e^{-\beta E_p} \; \la\m\n\la p \, 0 \, j'\n |\n| \widehat{T} |\n| p \, 0 \, j'\n \ra\m\n\ra^{\m *}}{\int_{-\infty}^{+\infty} \m {\rm d}p \; e^{-\beta E_p}} \right) \varrho_{jj'}^t\n \, \la j'\n | \mez .
      \end{eqnarray}
      Thus the sum of the third and the fourth traced lines of (\ref{derivative-hat-rho-take-2}) is equal to
      \begin{eqnarray} \label{piece-second+third-traced}
         - \, 2\,\pi \; {\cal N} \, \sum_{jj'\n} \; | j \ra \left( \frac{\int_{-\infty}^{+\infty} \m {\rm d}p \;\,
         e^{-\beta E_p} \; i \, \Bigl( \la\m\n\la p \, 0 \, j |\n| \widehat{T} |\n| p \, 0 \, j \ra\m\n\ra \, - \,
         \la\m\n\la p \, 0 \, j'\n |\n| \widehat{T} |\n| p \, 0 \, j'\n \ra\m\n\ra^{\m *} \Bigr)}
         {\int_{-\infty}^{+\infty} \m {\rm d}p \; e^{-\beta E_p}} \right) \varrho_{jj'}^t\n \, \la j'\n | \mez .
      \end{eqnarray}

      \mez Let us also perform the trace in the last line of (\ref{derivative-hat-rho-take-2}).
      Here we provide two hints for the sake of clarity:
      Firstly, it turns out that nonzero contributions may arise only if
      $(\nu''\n=\nu'''\n,j=j'\n,j'''\n=j''\n)$ or $(\nu''\n=\nu'''\n,j=j''\n,j'''\m=j'\n)$.
      Secondly, under such circumstances we have
      \be
         \delta(p'''\n-p''\n) \; \delta_{\rm energy}\m\Bigl(E_{p'''}+E^{\rm QE}_{j'''0}+\nu'''\n\hbar\omega-E_{p''}-E^{\rm QE}_{j''0}-\nu''\n\hbar\omega+E^{\rm QE}_{j0}-E^{\rm QE}_{j'0}\Bigr)
         \; = \; \frac{m}{|p''|} \, \delta(p=0) \, \delta(p''-p''') \mz . \mz
      \ee
      Keeping in mind these hints, one gets the following outcome:
      \begin{eqnarray}
         \hspace*{-1.50cm} & + & 4\,\pi^2 \, {\cal N} \, \sum_{j''} \, | j'' \ra \left\{
         \sum_{j} \sum_{\nu''} \sum_{\eta=\pm1} \, \frac{
         \int_{-\infty}^{+\infty} \m \frac{m\,{\rm d}p''}{\sqrt{2\,m\,E_{p''}}} \,
         \Theta(E_{p''}+E^{\rm QE}_{j''0}+\nu''\n\hbar\omega-E^{\rm QE}_{j0}) \, e^{-\beta E_p} \,
         \Bigl| \, \la\m\n\la p''\n \, \nu''\n \, j''\n \, |\n| \widehat{T} |\n| (\eta\, p) \, 0 \, j \ra\m\n\ra \, \Bigr|^2
         }{\int_{-\infty}^{+\infty} \m {\rm d}p' \; e^{-\beta E_{p'}}} \, \varrho_{jj}^t
         \right\} \la j'' | \nonumber\\
         \hspace*{-1.50cm} &  & \\
         \hspace*{-1.50cm} & + & 4\,\pi^2 \, {\cal N} \, \sum_j \sum_{j'\n \neq j} \, | j \ra \left\{ \sum_{\nu''} \sum_{\eta=\pm1} \,
         \frac{
         \int_{-\infty}^{+\infty} \m \frac{m\,{\rm d}p''}{\sqrt{2\,m\,E_{p''}}} \, \Theta(E_{p''}+\nu''\n\hbar\omega) \, e^{-\beta E_p}
         }{\int_{-\infty}^{+\infty} \m {\rm d}p' \; e^{-\beta E_{p'}}} \,
         \la\m\n\la p''\n \, \nu''\n \, j\n \, |\n| \widehat{T} |\n| (\eta\, p) \, 0 \, j \ra\m\n\ra \;
         \la\m\n\la p''\n \, \nu''\n \, j'\n \, |\n| \widehat{T} |\n| (\eta\, p) \, 0 \, j'\n \ra\m\n\ra^*
         \varrho_{jj'}^t\n \right\} \la j' | \; ; \nonumber
      \end{eqnarray}
      where $p>0$ is fixed as usual by the on-shell energy conservation condition.

\mez Putting all the ingredients together, instead of (\ref{derivative-hat-rho-take-2}) we have now
      \begin{eqnarray} \label{master-equation-take-1}
         \hspace*{-1.50cm} & & \frac{{\rm d}}{{\rm d}t} \, \varrho_{\rm S}^t \; = \nonumber\\
         \hspace*{-1.50cm} & = & \sum_j
         \sum_{j' \neq j} \; | j \ra \left(-\frac{i}{\hbar}\right)\m(E^{\rm QE}_{j0}-E^{\rm QE}_{j'0}\n) \, \varrho_{jj'}^t\n \, \la j'\n | \\
         \hspace*{-1.50cm} & - & \sum_j
         \sum_{j' \neq j} \; 2\,\pi \; {\cal N} \, \sum_{jj'\n} \; | j \ra \left( \frac{\int_{-\infty}^{+\infty} \m {\rm d}p \;\,
         e^{-\beta E_p} \; i \, \Bigl( \la\m\n\la p \, 0 \, j |\n| \widehat{T} |\n| p \, 0 \, j \ra\m\n\ra \, - \,
         \la\m\n\la p \, 0 \, j'\n |\n| \widehat{T} |\n| p \, 0 \, j'\n \ra\m\n\ra^{\m *} \Bigr)}
         {\int_{-\infty}^{+\infty} \m {\rm d}p \; e^{-\beta E_p}} \right) \varrho_{jj'}^t\n \, \la j'\n | \nonumber\\
         \hspace*{-1.50cm} & + & \sum_j \sum_{j'\n \neq j} \, 4\,\pi^2 \, {\cal N} \, | j \ra \left\{ \sum_{\nu''} \sum_{\eta=\pm1} \,
         \frac{
         \int_{-\infty}^{+\infty} \m \frac{m\,{\rm d}p''}{\sqrt{2\,m\,E_{p''}}} \, \Theta(E_{p''}+\nu''\n\hbar\omega) \, e^{-\beta E_p}
         }{\int_{-\infty}^{+\infty} \m {\rm d}p' \; e^{-\beta E_{p'}}} \,
         \la\m\n\la p''\n \, \nu''\n \, j\n \, |\n| \widehat{T} |\n| (\eta\, p) \, 0 \, j \ra\m\n\ra \;
         \la\m\n\la p''\n \, \nu''\n \, j'\n \, |\n| \widehat{T} |\n| (\eta\, p) \, 0 \, j'\n \ra\m\n\ra^*
         \varrho_{jj'}^t\n \right\} \la j' | \nonumber\\
         \hspace*{-1.50cm} & - & 2\,\pi \; {\cal N} \, \sum_{j} \; | j \ra \left( \frac{\int_{-\infty}^{+\infty} \m {\rm d}p \;\,
         e^{-\beta E_p} \; i \, \Bigl( \la\m\n\la p \, 0 \, j |\n| \widehat{T} |\n| p \, 0 \, j \ra\m\n\ra \, - \,
         \la\m\n\la p \, 0 \, j\n |\n| \widehat{T} |\n| p \, 0 \, j\n \ra\m\n\ra^{\m *} \Bigr)}
         {\int_{-\infty}^{+\infty} \m {\rm d}p \; e^{-\beta E_p}} \right) \varrho_{jj}^t\n \, \la j\n | \nonumber\\
         \hspace*{-1.50cm} & + & 4\,\pi^2 \, {\cal N} \, \sum_{j''} \, | j'' \ra \left\{
         \sum_{j} \sum_{\nu''} \sum_{\eta=\pm1} \, \frac{
         \int_{-\infty}^{+\infty} \m \frac{m\,{\rm d}p''}{\sqrt{2\,m\,E_{p''}}} \,
         \Theta(E_{p''}+E^{\rm QE}_{j''0}+\nu''\n\hbar\omega-E^{\rm QE}_{j0}) \, e^{-\beta E_p} \,
         \Bigl| \, \la\m\n\la p''\n \, \nu''\n \, j''\n \, |\n| \widehat{T} |\n| (\eta\, p) \, 0 \, j \ra\m\n\ra \, \Bigr|^2
         }{\int_{-\infty}^{+\infty} \m {\rm d}p' \; e^{-\beta E_{p'}}} \, \varrho_{jj}^t
         \right\} \la j'' | \; . \nonumber
      \end{eqnarray}
      The just derived formula (\ref{master-equation-take-1}) can already be interpreted as the sought master equation for
      $\hat{\varrho}_{\rm S}^t$. Observe that the evolution of diagonal populations $\varrho_{jj}^t$ is decoupled from coherences
      $\varrho_{jj'}^t$ $(j'\n \neq j)$. 

\subsection{Evolution of the populations}

\mez Proceeding further, we take equation (\ref{master-equation-take-1}) and extract for a given $j$ the temporal evolution
of $\varrho_{jj}^t$. One gets a much simpler evolution equation of the form
      \begin{eqnarray} \label{master-equation-populations-take-1}
         \hspace*{-1.50cm} & & \frac{{\rm d}}{{\rm d}t} \, \varrho_{jj}^t \; = \\
         \hspace*{-1.50cm} & = & - 2\,\pi \; {\cal N} \, \sqrt{\frac{2\,\pi\,m}{\beta}} \left(\,\int_{-\infty}^{+\infty} \m {\rm d}p \;\,
         e^{-\beta E_p} \; i \, \Bigl( \la\m\n\la p \, 0 \, j |\n| \widehat{T} |\n| p \, 0 \, j \ra\m\n\ra \, - \,
         \la\m\n\la p \, 0 \, j\n |\n| \widehat{T} |\n| p \, 0 \, j\n \ra\m\n\ra^{\m *} \Bigr) \right) \varrho_{jj}^t\n \nonumber\\
         \hspace*{-1.50cm} & + &
         \sum_{j'} \left( \sum_{\nu''} 4\,\pi^2 \, {\cal N} \, \sqrt{\frac{2\,\pi\,m}{\beta}} \, \sum_{\eta=\pm1} \,
         \int_{-\infty}^{+\infty} \m \frac{m\,{\rm d}p''}{\sqrt{2\,m\,E_{p''}}} \,
         \Theta(E_{p''}+E^{\rm QE}_{j0}+\nu''\n\hbar\omega-E^{\rm QE}_{j'\n0}) \, e^{-\beta E_p} \,
         \Bigl| \, \la\m\n\la p''\n \, \nu''\n \, j\n \, |\n| \widehat{T} |\n| (\eta\, p) \, 0 \, j'\n \ra\m\n\ra \, \Bigr|^2 \right)
         \varrho_{j'\n j'}^t \mez . \nonumber
      \end{eqnarray}
      We have tacitly incorporated here an obvious property
      \be
         \int_{-\infty}^{+\infty} \m {\rm d}p \; e^{-\beta E_p} \; = \; \sqrt{\frac{2\,\pi\,m}{\beta}} \mez .
      \ee

      \mez Proceeding towards finalizing all our elaborations, we may simplify the second line of (\ref{master-equation-populations-take-1})
      by exploiting the unitarity property (\ref{unitarity-1-take-3}) from S2. This yields
      \begin{eqnarray} \label{master-equation-populations-take-2}
         \hspace*{-1.50cm} & & \frac{{\rm d}}{{\rm d}t} \, \varrho_{jj}^t \; = \\
         \hspace*{-1.50cm} & = & - \, \varrho_{jj}^t\n \, \sum_{j'}
         \left( \sum_{\nu''} 4\,\pi^2 \, {\cal N} \, \sqrt{\frac{2\,\pi\,m}{\beta}} \, \sum_{\eta=\pm1} \,
         \int_{-\infty}^{+\infty} \m \frac{m\,{\rm d}p''}{\sqrt{2\,m\,E_{p''}}} \,
         \Theta(E_{p''}+E^{\rm QE}_{j'0}+\nu''\n\hbar\omega-E^{\rm QE}_{j0}) \, e^{-\beta E_p} \,
         \Bigl| \, \la\m\n\la p''\n \, \nu''\n \, j'\n \, |\n| \widehat{T} |\n| (\eta\, p) \, 0 \, j \ra\m\n\ra \, \Bigr|^2 \right)
          \nonumber\\
         \hspace*{-1.50cm} & + &
         \sum_{j'} \left( \sum_{\nu''} 4\,\pi^2 \, {\cal N} \, \sqrt{\frac{2\,\pi\,m}{\beta}} \, \sum_{\eta=\pm1} \,
         \int_{-\infty}^{+\infty} \m \frac{m\,{\rm d}p''}{\sqrt{2\,m\,E_{p''}}} \,
         \Theta(E_{p''}+E^{\rm QE}_{j0}+\nu''\n\hbar\omega-E^{\rm QE}_{j'\n0}) \, e^{-\beta E_p} \,
         \Bigl| \, \la\m\n\la p''\n \, \nu''\n \, j\n \, |\n| \widehat{T} |\n| (\eta\, p) \, 0 \, j'\n \ra\m\n\ra \, \Bigr|^2 \right)
         \varrho_{j'\n j'}^t \mz ; \nonumber
      \end{eqnarray}
      where $p>0$ is again fixed by the energy conservation condition as appropriate. Formula (\ref{master-equation-populations-take-2})
      motivates us to introduce the Floquet transition rates
 \begin{eqnarray} 
 & & a_{j'\n j}^{\nu} \; = \; {\cal N} \, \mathcal{Z}^{-1} \m \int_{-\infty}^{+\infty} \m {\rm d}p
 \int_{-\infty}^{+\infty} \m {\rm d}p' \; e^{-\beta\frac{p^2}{2\,m}} \,
 \delta\m\m\left(\frac{p'^2}{2\,m}+E_{j'\n\nu}^{\rm QE}-\frac{p^2}{2\,m}-E_{j0}^{\rm QE}\right) \,
 \Bigl| \la\m\n\la p'\n\,\nu\,j'\n |\n| \widehat{T} |\n| p\,0\,j \ra\m\n\ra \Bigr|^2 \;\;\; ; \nonumber
 \end{eqnarray}
      in consonance with the previously given definition (\ref{Floquet-rates-Appendix-A}).
      We have conveniently absorbed here the factor $4\,\pi^2$ into ${\cal N}$.
      Then the second line of (\ref{master-equation-populations-take-1}) equals simply to
      \be
         - \; \varrho_{jj}^t \sum_{j'} \, a_{j'\n j} \mez ;
      \ee
      whereas the last line takes the form
      \be
         + \; \sum_{j'} \, a_{jj'} \, \varrho_{j'\n j'}^t \mez .
      \ee
      If so, then our master equation (\ref{master-equation-populations-take-1}) for the populations
      can be redisplayed in its finalized form
      \be \label{master-equation-populations-take-3}
         \frac{{\rm d}}{{\rm d}t} \, \varrho_{jj}^t \; = \; \sum_{j'} \, a_{jj'} \, \varrho_{j'\n j'}^t \; - \; \varrho_{jj}^t \sum_{j'} \, a_{j'\n j} \; ;
      \ee
which agrees with the Pauli rate equation (1) from the main text once one sets $\varrho_{jj}^t=\wp_j$. 


\end{document}